\documentclass[aps,pre,twocolumn,groupedaddress,showpacs]{revtex4-1}
\usepackage[utf8]{inputenc}
\usepackage[english]{babel}
\usepackage{amsmath,graphicx,enumerate,amssymb}
\usepackage{epstopdf}
\usepackage{multirow}
\usepackage{color}

\newcommand{\CID}{\text{CID}}

\begin{document}

\title{Vicsek Model by Time-Interlaced Compression:\\ a Dynamical Computable Information Density}
\author{A. Cavagna$^{1,2}$, P.M.Chaikin$^{3}$, D. Levine$^{4}$, S. Martiniani$^{5}$, A. Puglisi$^{1,2}$, M. Viale$^{1,2}$}
\affiliation{
	$^1$ Dipartimento di Fisica - Universit\`a La Sapienza - 00185 Rome, Italy \\
	$^2$ Istituto dei Sistemi Complessi - Consiglio Nazionale delle Ricerche, 00185 Rome, Italy \\
	$^3$ Center for Soft Matter Research, Department of Physics, New York University, 10003 New York, USA \\
	$^4$ Department of Physics, Technion-IIT, 32000 Haifa, Israel \\
	$^5$ Department of Chemical Engineering \& Materials Science, University Of Minnesota, 55455 Minneapolis, USA
	}
\date{\today}

\begin{abstract}
Collective behavior, both in real biological systems as well as in
theoretical models, often displays a rich combination of different
kinds of order. A clear-cut and unique definition of ``phase" based on
the standard concept of order parameter may therefore be complicated,
and made even trickier by the lack of thermodynamic equilibrium.
Compression-based entropies have been proved useful in recent years in
describing the different phases of out-of-equilibrium systems.  Here, we
investigate the performance of a compression-based entropy, namely the Computable Information Density (CID), within the
Vicsek model of collective motion. Our entropy is defined through a
crude coarse-graining of the particle positions, in which the key role
of velocities in the model only enters indirectly through the velocity-density coupling.  We discover that such entropy is a valid
tool in distinguishing the various noise regimes, including the
crossover between an aligned and misaligned phase of the velocities,
despite the fact that velocities are not used by this entropy.
Furthermore, we unveil the subtle role of the time coordinate,
unexplored in previous studies on the CID: a new
encoding recipe, where space and time locality are both preserved on
the same ground, is demonstrated to reduce the CID.  Such an improvement is particularly significant when working
with partial and/or corrupted data, as it is often the case in real biological experiments.
\end{abstract}

\pacs{}

\maketitle

\section{Introduction}
Statistical physics and information theory
have a long history of cross-fertilization~\cite{zurek2018complexity}, with both disciplines
relying on a key concept: a quantitative statistical measure of order~\cite{chaitin1990information,mezard2009information}.  
For physical systems both in and out of  equilibrium, such a measure is the starting point for a general theory
of phase transitions, and it is a prerequisite for the study of  response
 to external perturbations~\cite{BPRV08}. The
statistical notion of entropy - the fundamental link
between thermodynamics and equilibrium statistical mechanics - is
the main inspiration for the central concept 
of  information theory, the Shannon entropy~\cite{shannon}.  More recently,
information theoretic ideas have proven useful in the study of physical many-body 
systems, for example, to obtain good estimates for critical temperatures in equilibrium spin
systems~\cite{vogel2009phase,melchert}, 
to describe entropy production in
the context of stochastic thermodynamics~\cite{LS99,parrondo2009},
and to obtain accurate estimates of the entropy in both equilibrium and non-equilibrium 
systems~\cite{martiniani2019quantifying, martiniani2020correlation}.

In this paper, we seek a method
to measure order in some non-equilibrium flocking models, with an eye
to eventually analyzing observational data on living systems.  In particular, we propose a  
novel analysis which treats ordering in space and time on an equal footing.  We will argue
that by analyzing the temporal development of spatial patterns together, we can
obtain information in a way that is robust enough to survive 
the inevitable noise present in collections of living objects.

Our approach is based on a recently proposed information-based
measure of order for out-of-equilibrium systems~\cite{martiniani2019quantifying}.  This proposal, called {\em Computable
 Information Density} (CID), relates to the compression rate
measured by lossless compression algorithms~\cite{loreto,puglisi2003data}.   CID was shown to give
clear signatures of important transitions in the systems studied, which included
several absorbing state models~\cite{Non-Equilibrium_Book} as well as an active matter model,
repulsive active Brownian particles (ABP), where motility induced
phase separation appears at large enough
concentrations~\cite{cates15}.  We note that these models
lack first-principles Hamiltonians, which makes this
quantification of order even more compelling.

In particular, we compress a sequence $x$ using a universal lossless compression algorithm (such as one of the Lempel-Ziv algorithms)\cite{cover2012elements}, and define the CID of $x$ as
\begin{equation}
\text{CID} \equiv \frac{\mathcal{L}(x)}{N}.
\end{equation}
where $\mathcal{L}(x)$ is the total binary code length of the compressed sequence, and $N$ is the length of the original sequence $x$.  
We note that although we are using the term ``sequence'', we are not restricted to 1D strings -- microstates in any dimension may be 
compressed by appropriate procedures \cite{sheinwald1990two}, as will be discussed later.  For equilibrium systems, the CID gives a good
approximation to the thermodynamic entropy \cite{martiniani2019quantifying}.

Here, we investigate the effectiveness of CID
in flocking models, a class of active matter known to exhibit a
different kind of order from ABP.  We introduce a novel ``space-time''
algorithm based on CID
which is sensitive enough to yield insight even when it only uses strongly coarse-grained
information about the system, such as a binarized
(empty/occupied) density field.  
Our main results are:
\begin{itemize}
\item We are able to accurately characterize  phase behavior by CID using a coarse-grained representation that only considers occupied/unoccupied discrete regions of space.
\item A single parameter Q, characterizing the amount of correlations in the system as measured by CID \cite{martiniani2020correlation}, can accurately detect multiple phases in the system.
\item By proper incorporation of temporal as well as spatial information, we are able to capture details of phase transitions with
greater sensitivity.
\item This increase in sensitivity allows us to measure important features even when our data is substantially corrupted, such as is typically the case for
data from actual living systems.
\end{itemize}
In Section II we
introduce the flocking model used to test our
approach, the two-dimensional Vicsek Model, following which
we explain the encoding method we use to translate its dynamical
configurations into a binary string, and we then define our order measure. In Section III we present 
results  obtained from numerical simulations, and we present conclusions
in Section IV.

\section{Encoding Dynamical Configurations}

\subsection{The Vicsek Model}

\begin{figure}[b!]
	\includegraphics[width=\columnwidth]{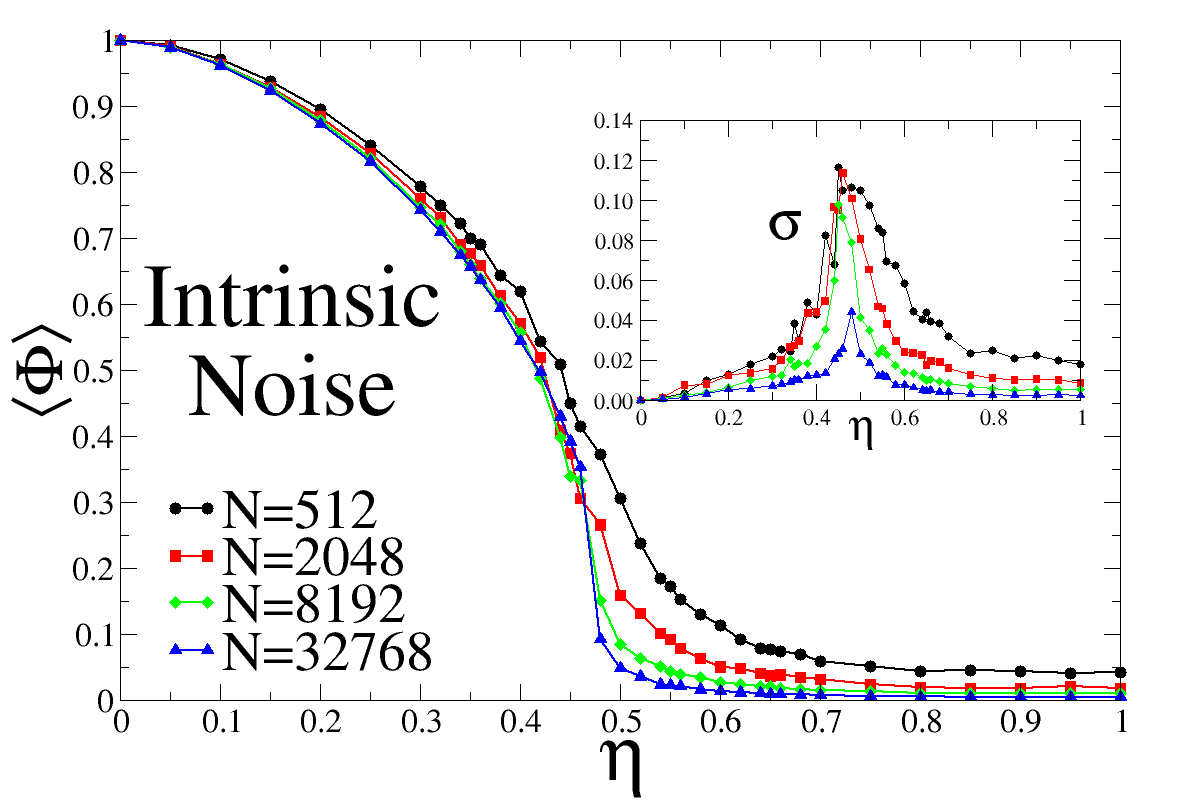}
	\includegraphics[width=\columnwidth]{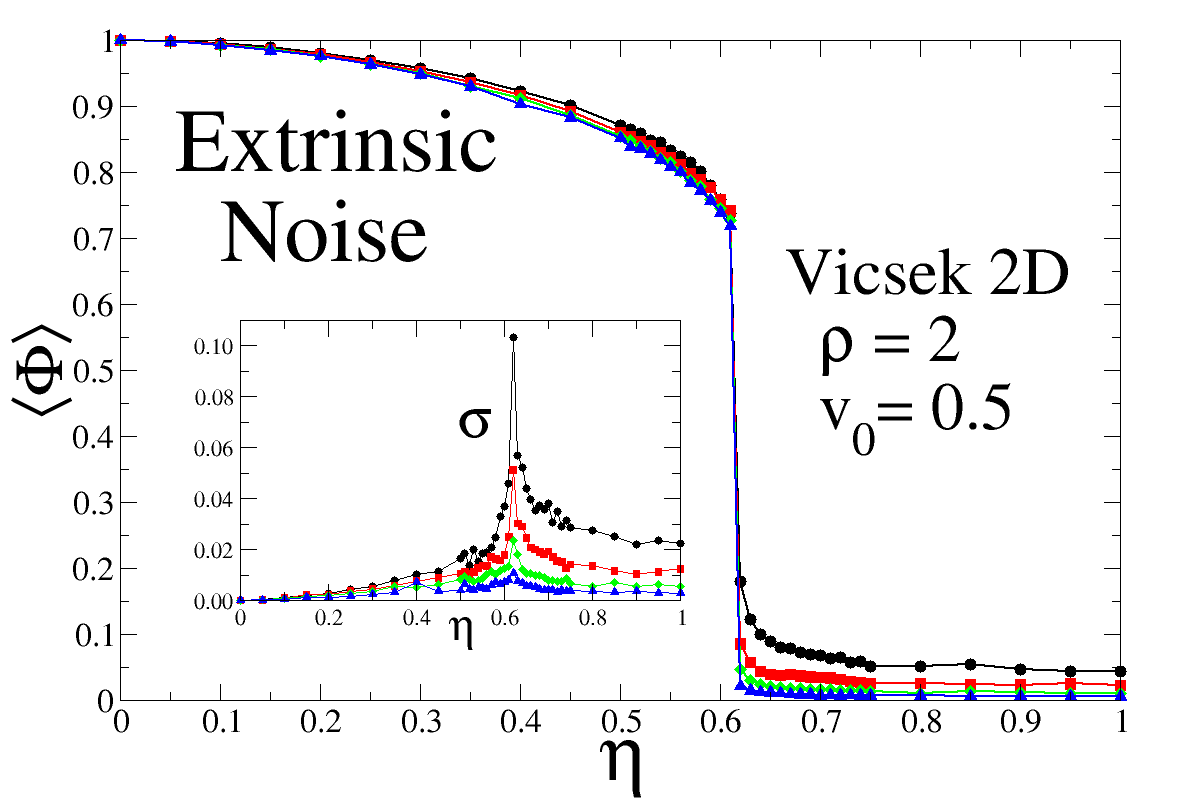}
	\caption{Mean $\langle \Phi \rangle$ of the order parameter
          $\Phi=\left|\sum_i e^{i\theta_i}/N\right|$ and its
          fluctuations $\sigma^2=\langle(\Phi - \langle \Phi
          \rangle)^2 \rangle$ for several system sizes $N$, 
          as a function of noise strength $\eta$, for intrinsic (top) and extrinsic (bottom) cases.
	\label{fig:Vicsek}}
\end{figure}
We will consider here an archetypical model of collective behavior in
biological systems, the Vicsek model (VM)~\cite{vicsek+al_95}, which we will study in two dimensions.  The VM is an active matter model in which each
agent/particle updates its velocity by imitating the average velocity
of its nearest neighbors, and then updates its position by following its
own velocity \cite{parrish_review,gregoire2003moving}. The first is an
alignment mechanism, identical to classical ferromagnetic models with
continuous symmetry, while the second leads to diffusion-like motion. It is worth noting that the interaction (or connectivity)
matrix $n_{ij}(t)$ depends on time through the inter-particle
distances $r_{ij}(t)$, which in turn depend on the velocities; this closes a feedback loop between positions and velocities
that drives the systems out-of-equilibrium. As a function of noise strength,
the VM displays a (finite-size) transition between a
high-noise disordered phase and a low-noise ordered (or polarized)
phase. Close to this transition, non-trivial fluctuations
emerge both in velocity and density, whose properties depend on the
nature of the noise and interaction
\cite{aldana2007phase,baglietto2009nature}.  

More specifically,  the model consists of a system of $N$
active particles, each moving in a 2D plane with fixed speed $v_0$ in a square box of
area $A= L \times L $ with periodic boundary conditions.  At each
time step the particles tend to align their direction of motion with that of
their neighbors, with some noise added to make the dynamics stochastic.  Let ${\bf
  r}_i(t)\equiv\left(x_i(t),y_i(t)\right)$ and $\theta_i(t)$ be the position
and  orientation respectively of particle $i$ at some time $t$, $\mathcal{N}_i(t)$ the set of its neighbors in a
circle of radius $R$ \footnote{hence this is a {\it metric} implementation of VM} centered about the particle $i$, and ${\bf
  V_i}(t) \equiv \sum_{j \in \mathcal{N}_i(t) } {\bf v_j}(t) / |\mathcal{N}_i(t)|$ the average
velocity of the particles in the neighborhood of $i$.  The
velocity vector will be represented in the complex plane as ${\bf v}_i(t)
= v_0 e^{i\theta_i(t)}$.  

There are two cases to consider: {\it intrinsic} (also called {\it vectorial}) and {\it extrinsic} (also called {\it scalar}) noise.  
For the case of intrinsic noise, 
each particle in the system evolves
according to the update rule
\begin{eqnarray*}
\theta_i(t+1) &=& \Theta\left[ {\bf V}_i(t) \right] + \eta\; \xi_i(t) \\
{\bf r}_i(t+1) &=& {\bf r}_i(t) + {\bf v}_i(t+1)
\end{eqnarray*}
where $\Theta\left[ {\bf v} \right] \equiv \text{atan2}(v_y,v_x)$ is the angular coordinate of ${\bf v}$ ,
$\xi_i(t)$ is a uniformly distributed random variable in $[-\pi, \pi]$ and $\eta \in [0,1]$ is the noise strength.
For extrinsic noise, the updating rule for $\theta_i$ is a little different:
$$
\theta_i(t+1) = \Theta\left[ {\bf V}_i(t) + \eta e^{i\xi_i(t)}\right] \\
$$

Both types of noise generate a phase diagram characterized by two critical
noise values $\eta_b<\eta_c$ which, finite size effects apart, depend
on the density $\rho$ of the system and the speed $v_0$ of the
particles.  When the noise intensity is large enough, $\eta > \eta_c$, a
Vicsek fluid stays in a fully disordered state with a spatially
homogeneous density and a thermal-like distribution of particle
directions.
By reducing the noise to $\eta \simeq \eta_c$ the
rotational symmetry is spontaneously broken and the system exhibits
collective motion: a clear polarization transition appears for the
velocities, which is accompanied by strong modifications of the
density field, similar to a phase separation with traveling band
formation \cite{solon}.   

The chief difference between the two kinds 
of noise is that extrinsic noise has a stronger contribution to band formation 
than intrinsic noise.  In the latter case, the traveling bands are less clear
and defined, though they become more visible
as the size of the system increases.  When the noise is further
reduced to $\eta<\eta_b$, the bands disappear (sharply with extrinsic
noise, smoothly with intrinsic noise), replaced by a polarized state
whose density is spatially homogeneous but possessing giant
fluctuations \cite{ginelli}.  At finite $N$, a
sharp transition for extrinsic noise
\cite{gregoire+al_04,chate+al_08b}, and a smoother transition for 
intrinsic noise \cite{nagyvicsek} are usually found.  The order of the transition in
the large $N$ limit is still under debate, as
is whether or not the nature of the intermediate phase of traveling bands is
an artifact of the periodic boundary conditions
\cite{aldanadebate}. In this work we will not discuss the
asymptotic nature of the transition, but just show how we can
distinguish the phases of VM through a single observable related to
the CID of a suitable coding of system configurations.  In
Figure \ref{fig:Vicsek} we show the polarization $\langle \Phi \rangle$ and its fluctuations for
the simulated systems, where
$$
\Phi=\frac{1}{N}\left|\sum_i e^{i\theta_i}\right|
$$
indicating the nature of the transitions for the two types of noise.

\subsection{Coarse-Graining and Alphabet of Symbols}

To implement the CID measurements, we first discretize our space by 
overlaying a regular square grid of $M=m \times m$ cells of size $b=L/m$, 
and we observe the system evolution for a time window of $T$ steps.
At each time $t$ we assign the symbol `1' to all the cells occupied by
at least one particle and the symbol `0' to empty ones: in this
way we build a 3-dimensional array of $M\times T$ bits which we denote ${\cal {A}}$.  

We note that we could have considered a richer alphabet, e.g. based upon combinations of
local density and local (e.g. cell averaged) velocity, but there are 
good reasons for not doing this.  First, in real biological data, we typically have direct access
only to the positions of the agents; other degrees of freedom
(e.g. velocities) are obtained from the knowledge of the positions.
Since we will be simultaneously encoding $T>1$ configurations, 
we expect  the velocity information to be present implicitly.

\subsection{Scanning: The Z-Order Curve}
\begin{figure}[t!] 
  \includegraphics[width=\columnwidth]{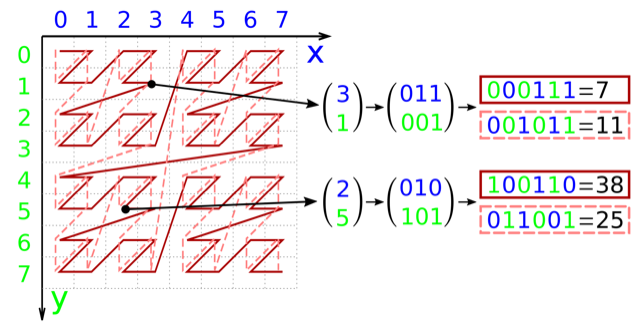} 
  \caption{ Procedure to get the Z-value, \textit{i.e.} the
    coordinate along the Z-order curve which is a one-dimensional
    spanning of a multi-dimensional space, preserving locality. Each point has a Z-value which is obtained by
    interleaving the bits of the $x$ coordinate with the bits of the
    $y$ coordinate. This, in two dimensions, can be done in two
    different ways: bits can be interleaved in the order $xyxyxy\dots$ ($x$-first) or in the order $yxyxyx\dots$ ($y$-first). The two possibilities are
    illustrated for two different points, using colors (green and
    blue) to make appreciable the choice of the interleaving
    order. One choice gives place to the solid curve, the other gives
    place to the dashed curve.  In the example, the distance (along
    the Z-curve) between the two points depends upon the choice of
    the order: it is $38-7=31$ for $x$-first order, and $25-11=14$ for $y$-first order.}
\label{fig:zCurve}
\end{figure}

Since the typical compression programs operate on one-dimensional strings of characters, 
we need to scan the array ${\cal {A}}$ and produce a 
1D sequence. Different scanning procedures exist, but in this work we will employ two procedures
based on the so-called 
{\it Z-order} or {\it Morton-order} mapping \cite{morton1966computer}, which is similar to Hilbert scanning \cite{sheinwald1990two}.  This class of  mappings have the advantages of preserving spatial locality in a reasonable fashion, and of working in 
arbitrary dimensions.  We will discuss the \textit{Z-order} mapping for the 2D case; generalization to higher
dimensions is immediate.

In 2D, a state of the system is represented by a matrix $M$ with entries $M_{mn}$ \footnote {In our case,
the entries take on the values $0$ or $1$, but this is not important for the scan.}.
We wish to compose a 1D string $\sigma$ with entries $\sigma_{l}$ which are derived from
$M$. The Z-order algorithm does this as follows:
\begin{enumerate}
\item Write the integers $n$ and $m$ in binary representation, such
that $n_{k}$ and $m_{k}$ are the $k^{th}$ digits in the representations.
\item Interleave the digits of the $n$ and $m$ to form a new binary string
$n_{1}m_{1}n_{2}m_{2}n_{3}m_{3}...$.  If the binary string for $n$ or $m$ 
is shorter than the other, pad it with zeros.   This is the binary representation of
some integer $l$.
\item Set $\sigma_{l} = M_{mn}$.
\end{enumerate} 
We note that there are two such possible mappings, (the interlaces $n_{1}m_{1}n_{2}m_{2}n_{3}m_{3}...$
and $m_{1}n_{1}m_{2}n_{2}m_{3}n_{3}...$) as seen in Figure \ref{fig:zCurve}. 
We will discuss this ambiguity later.
The generalization to higher dimensions is straightforward: the binary representations
of the lattice site $(j,k,...,r)$ are interleaved in the same fashion as above.
 
\begin{figure}[b!]
  \includegraphics[width=1\columnwidth]{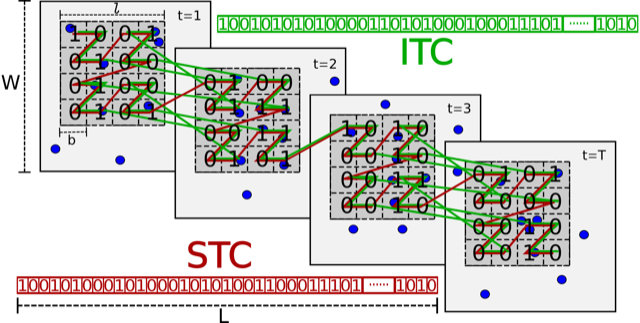}
  \caption{Comparing STC (red curve) and ITC (green curve) schemes.
  STC simply places the configurations one behind the other after a two-dimensional Z-order curve cells spanning over the cells of a single configuration.
  Differently, ITC interlaces bits of configurations at different times following a three-dimensional Z-order curve.
  Note: the number of $1$ and $0$ are identical, only the positions in the output one-dimensional string are different.}
  \label{fig:Coding}
\end{figure}

In this work, we wish to study a sequence of $T>1$ sequential configurations showing the time evolution
of the system.  There are two main ways to do this (see Fig.~\ref{fig:Coding}):
\begin{itemize}
\item{\bf Serialized Time Coding (STC)}
We scan the 2D configuration matrix M at each time step using the 2D Z-order algorithm, and we then 
concatenate the resultant 1D strings in a sequence of length  $m \times m \times T$ according to their time order.
\item{\bf Interlaced Time Coding (ITC)}
We scan the entire space-time array ${\cal {A}}$ with the 3D Z-order algorithm, producing a 1D sequence of length  $m \times m \times T$.  This method
is expected to better preserve time correlations.
\end{itemize}

\twocolumngrid

We will show that ITC is more sensitive to the ordering than STC, which we argue will play a useful role in analyzing collective dynamics of living things.

For our analysis, we find that it is useful to consider a quantity $Q$ \cite{martiniani2020correlation} defined by
$$
Q(x)=1-\frac{\CID(x)}{\langle {\CID(x_\text{sh})}\rangle}
$$
where $\CID(x_\text{sh})$ is the CID of a random shuffle of the sequence $x$,
and $\langle {\CID(x_\text{sh})}\rangle$ indicates an average over several such
shuffled sequences.  For asymptotically long strings, $0 \le Q \le 1$, with $Q \simeq 0$
indicating that the data string is uncorrelated, while $Q>0$ indicates the presence of
order.  
 
We note one additional technical detail.  The Z-curve does not span the space (or the space-time for ITC
case) in an isotropic way, but we can improve the isotropy by averaging 
$Q$ over the different interleavings of coordinates as shown in Figure \ref{fig:zCurve}.  
For STC, there are 2 possible interleavings (as shown in Figure \ref{fig:zCurve}) and
for ITC there are $3!=6$ interleavings, since the Z-curve is taken in the full 3D matrix.  
We define $Q_I$ and $Q_S$ as the average of $Q$:
$$
Q_{I,S} = \langle Q(x) \rangle_{\mathrm{int}}
$$
where the average is taken over the 2 (for ITC) or 6 (for STC) possible curves.

\section{Results}

In this Section we show how the dynamical information arises by an
analysis of $Q$ by comparing the ITC and STC schemes on simulations of the
2D Vicsek Model.  For both intrinsic and
extrinsic types of noise, we focus on a typical set of parameters, setting the
interaction radius, density and speed to $R=1$,
$\rho=N/V=2$ and $v_0=0.5$, respectively.  We simulate the model for different
sizes, from $N=2^9$ to $N=2^{15}$, and vary the noise strength
$0 \le \eta \le 1$.  The system size is $L \times L$, and we employ periodic
boundary conditions.

We choose our observation window (that is, the data which we analyze) to
be smaller than the entire system to avoid points which are near one another
in space being far apart in the Z-order curve.  Since such issues arise only near
the boundary, we choose an observation
window of size $L/2 \times L/2$ at most.
Finally, we average  $Q$  over many configurations ($10^3 -
10^4$).  We begin to collect configuration data for our
analysis after waiting for the system to reach a stationary
steady state, a time $t \simeq 10^4$.

\begin{figure}[h!] 
  \includegraphics[width=\columnwidth]{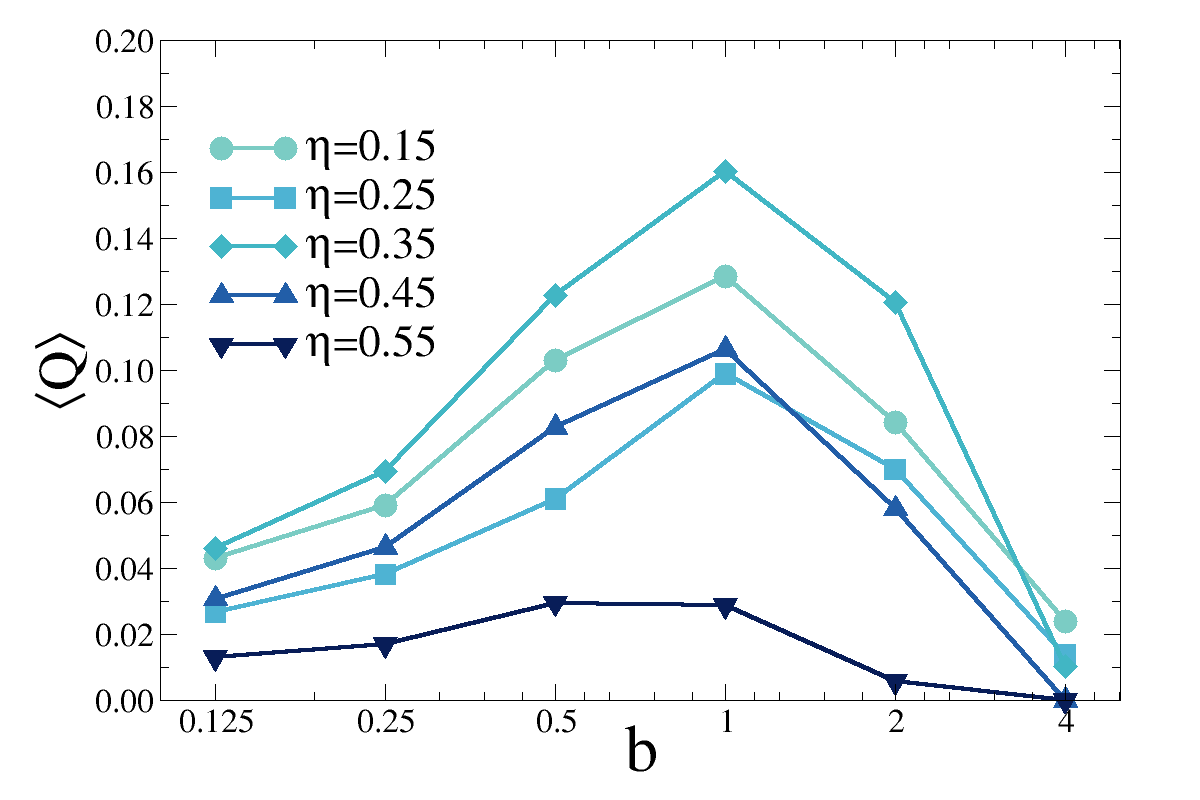}
\caption{The trend of $Q$ for $T=1$ by varying the cell size $b$.
	 As we expected, $Q$ tends to trivial value $0$ for $b\to 0$ and $b\to L$, 
	 consequently there must be a maximum for some intermediate value, in this case $b=1$.
	 (VM simulations with intrinsic noise, parameters are $N=32768$ and $L=64$)}
\label{fig:QVsBOX}
\end{figure}
We first  studied the effect of the cell size $b$ at $T=1$ (for
which there is no difference between STC and ITC), see
Fig.~\ref{fig:QVsBOX} which shows an optimal value (best compression)
at $b=1$ for most of the choices of the noise intensity.  It does not
seem a coincidence that this optimal value coincides with the
interaction radius $R$. Since we want to study the effect of other
parameters, we fix $b=1$ in the remainder of this work.

\subsection{Dependence On The Time Window $T$ And On The Time Encoding}

\begin{figure}[h!]
\includegraphics[width=\columnwidth]{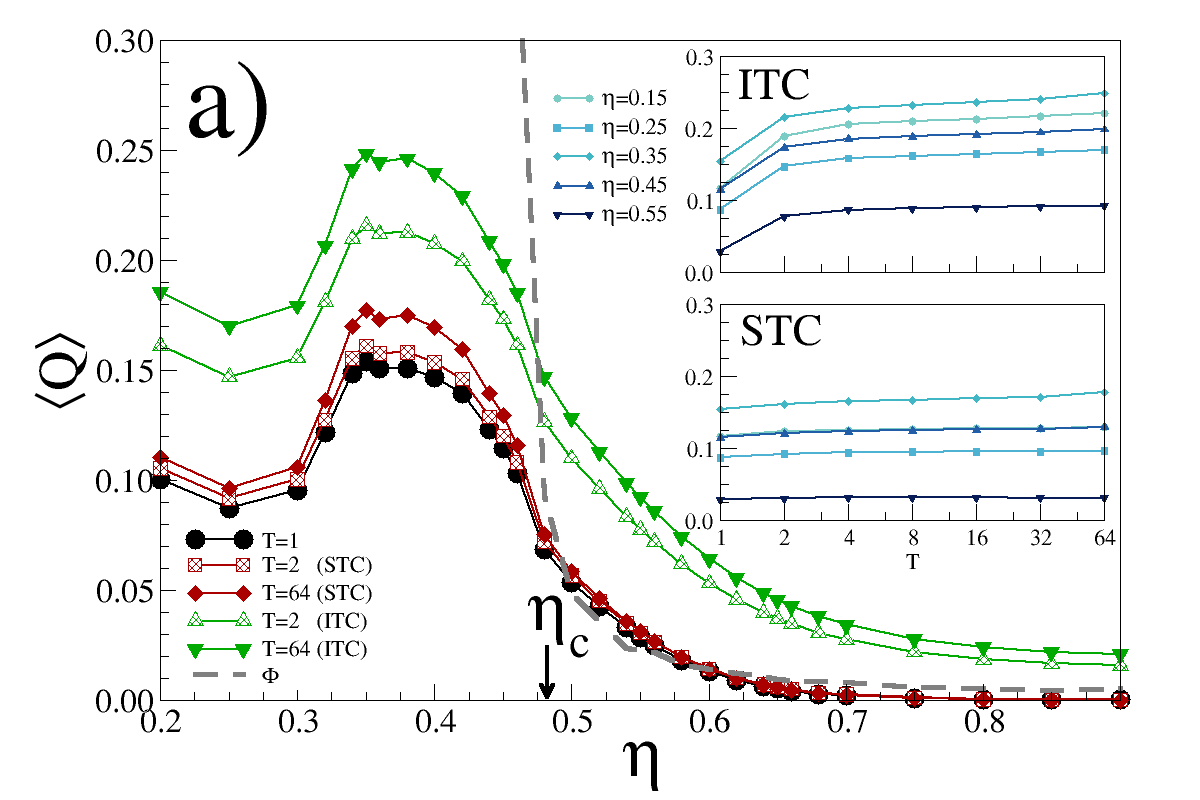}
\includegraphics[width=\columnwidth]{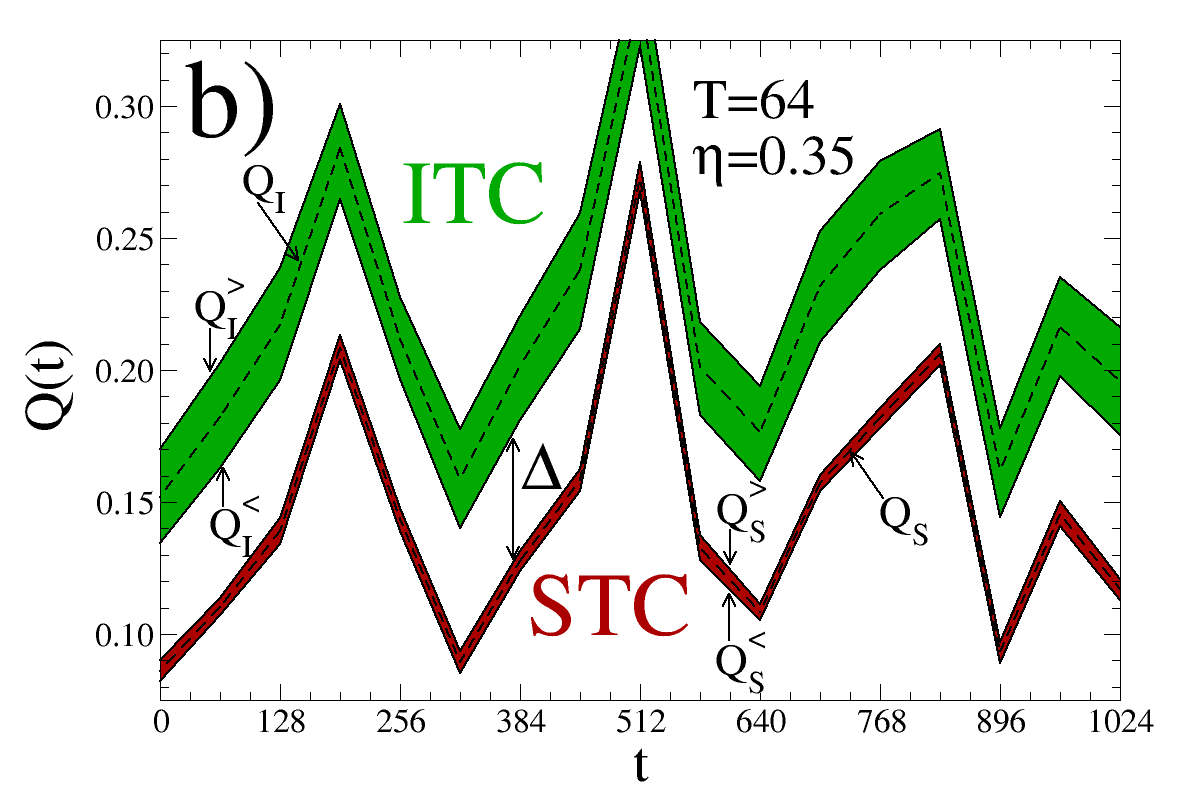}
\caption{\textbf{(a)} $Q$ computed by the simple time-concatenation STC
  scheme (red lines) and by the Z-order in space-time with ITC scheme
  (green lines) compared to $Q$ computed on a single configuration
  (black line $T=1$).  ITC scheme gives a $Q$ higher than STC
  already at $T=2$ and its $Q$ increases significantly faster than STC as $T\to 64$ (insets).\\ \textbf{(b)} Evolution of $Q$
  variability range due to permutations.  We show that STC and ITC
  schemes produce, not only in average but step by step also, $Q$
  values always well separated.  (VM simulations with intrinsic noise,
  parameters are $N=32768$ and $L=64$).}
\label{fig:QVsT}
\end{figure}

In Fig.~\ref{fig:QVsT}a we plot $\langle Q \rangle$ as a function of the noise $\eta$ for the ITC and STC protocols for the model
with intrinsic noise and different time-window lengths T.  Although the curves all have the same general shape, the dynamic range (maximum value - minimum value) of the
ITC protocol is considerably larger than that of the STC, suggesting that it might be a more sensitive measure of the order.  Fig.~\ref{fig:QVsT}b shows that the
value of $Q(t)$ is always higher for ITC than for STC, which is another indication of its better sensitivity. 
The main polarization transition at $\eta_c \simeq 0.48$ is seen as an 
inflection point in  $\langle Q \rangle$, which appears more cleanly in
the curves for the ITC scheme.
The polarization phase transition at $\eta_{c}$ is typically thought of as an ordering of the particle velocities.
For this reason, the fact that $Q$ with $T=1$ shows a clear signature of this transition is interesting.
For $T=1$, $Q$ is only a single spatial configuration from which   
no velocity information can be inferred.  This is a clear indication that when the velocities order at $\eta_{c}$, 
there is a simultaneous ordering in the density field, which cannot be appreciated by looking just at polarization.
Thus, the behavior of $Q$ is more sensitive
than the behavior of the polarization. In Section III.C we give details about how $Q$ describes the
full complex phase diagram of the VM. 
Here we focus on the effect of
$T$ and the way time is treated by the different CID encoding protocols.
When $T$ is increased, the shape of the curve $Q$ \textit{vs.} $\eta$ remains
 similar. The values of $Q$ obtained with the STC scheme
increase slightly, mainly because of the larger statistics of substrings. 
However, a significant increase of $Q$ is apparent when using the
ITC protocol.  For instance the ITC with $T=2$ is everywhere larger than 
STC with $T=64$, indicating a vastly greater sensitivity, even when
$\eta \to 1$, where the particles are evenly distributed with no
polarization.  The STC scheme preserves only space locality; therefore
at large noise values, where there are no spatial correlations, it
gives $Q\simeq 0$.  On the contrary the ITC scheme preserves space and time
locality and therefore exploits temporal correlations, resulting in $Q
> 0$ even at high noise.  Importantly, since temporal correlations are
large in the presence of significant order, the curve $\langle Q_I \rangle$
\textit{vs.} $\eta$ shows the polarization transition better and with higher resolution.

The difference between the two types of encoding is larger than
the statistical fluctuations, and thus is significant. 
In Fig. \ref{fig:QVsT}b we compare step-by-step the
``worst" result for ITC - the minimum value $Q^<_I$ over the 6
permutations - with the ``best" result for STC - the maximum value
$Q^>_S$ over the 2 permutations.  We find a gap $\Delta=Q^<_I - Q^>_S$
between the two encoding protocols which is always positive, a solid
test of the effective improvement achieved by ITC with respect to STC.

\begin{figure}[h!]
\includegraphics[width=\columnwidth]{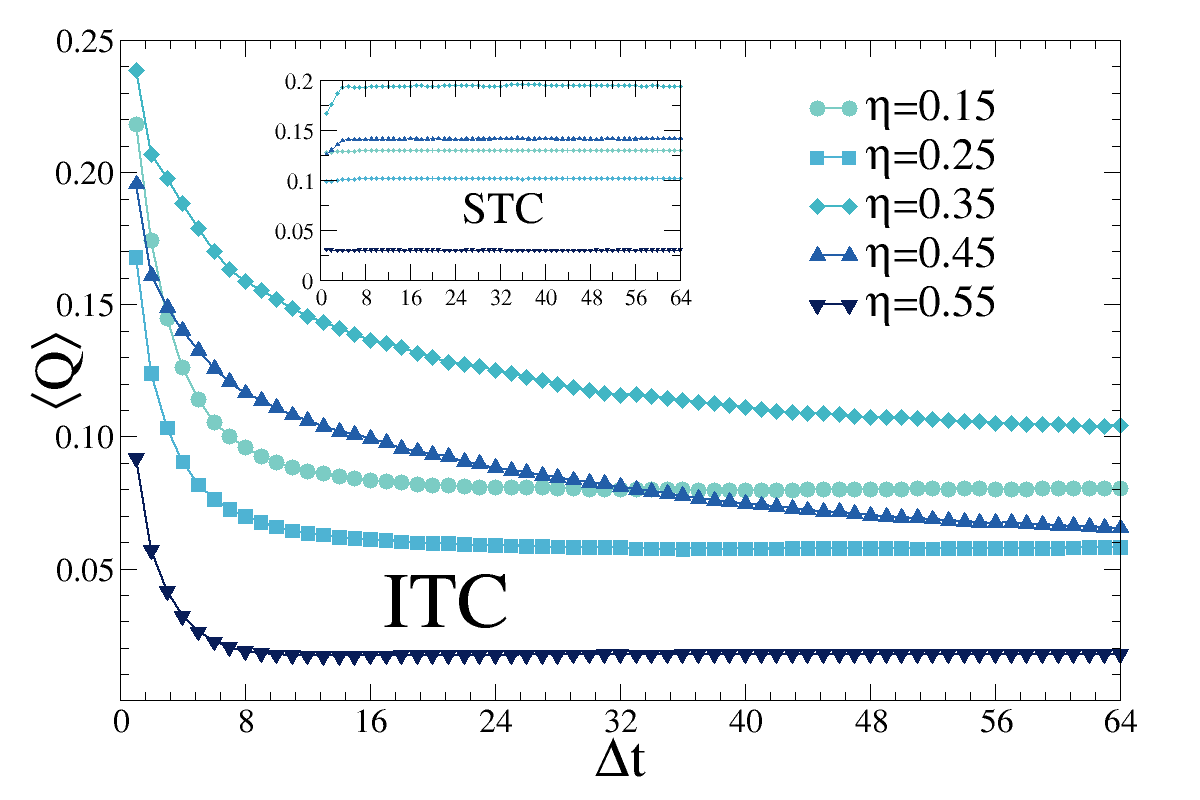}
\caption{Consequence on $Q$ when the sequences of $T$-length are built by
  skipping $\Delta t$ time steps.  ITC scheme shows a decay of $Q$ by
  varying $\Delta t$ that, as we expect, slows down as the transition
  is approached. Differently, the STC scheme (inset) does not
   show meaningful information.  (VM simulations with
  intrinsic noise, parameters are $N=32768$, $L=64$ and $T=64$).}
\label{fig:QVsDT}
\end{figure}

From these results we  conclude that temporal interlacing has added
information about dynamics, information that the STC scheme struggles
to reveal.  This is also appreciated by studying how $Q$ changes when
the distance $\Delta t$ between successive times in a sequence of $T$
configurations is increased (Fig.~\ref{fig:QVsDT}). This analysis
shows a striking difference between STC and ITC. STC is not
particularly sensitive to $\Delta t$:  it treats all
configurations as independent, without respect to their proximity in time. 
ITC, on the other hand, displays a smooth relaxation
towards an asymptotic value for large $\Delta t$.  The characteristic
time of this relaxation grows when $\eta \to \eta_c,\eta_b$ suggesting a power-law
rather than an exponential time decay, as in typical
slowing-down phenomena close to phase transitions.  

\subsection{Probing the VM Phase Diagram with $Q$}

\begin{figure}[h!] 
  \includegraphics[width=0.49\columnwidth]{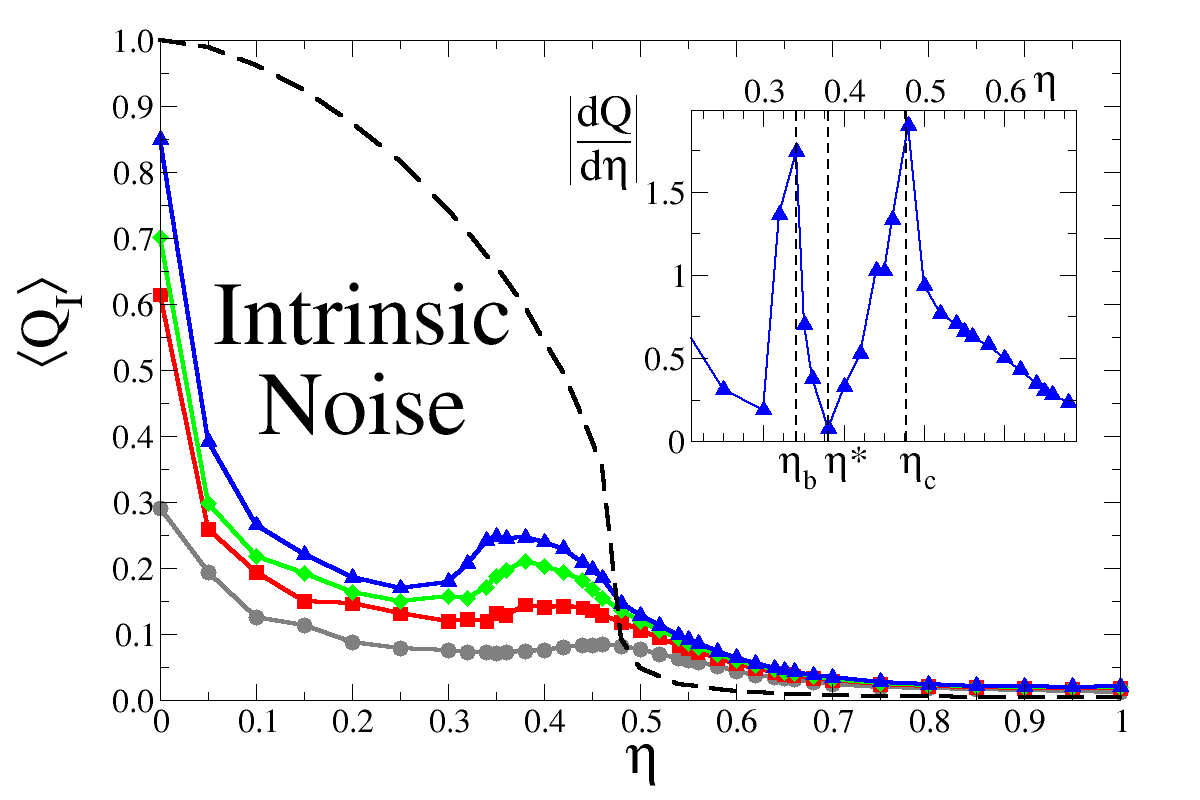} 
  \includegraphics[width=0.49\columnwidth]{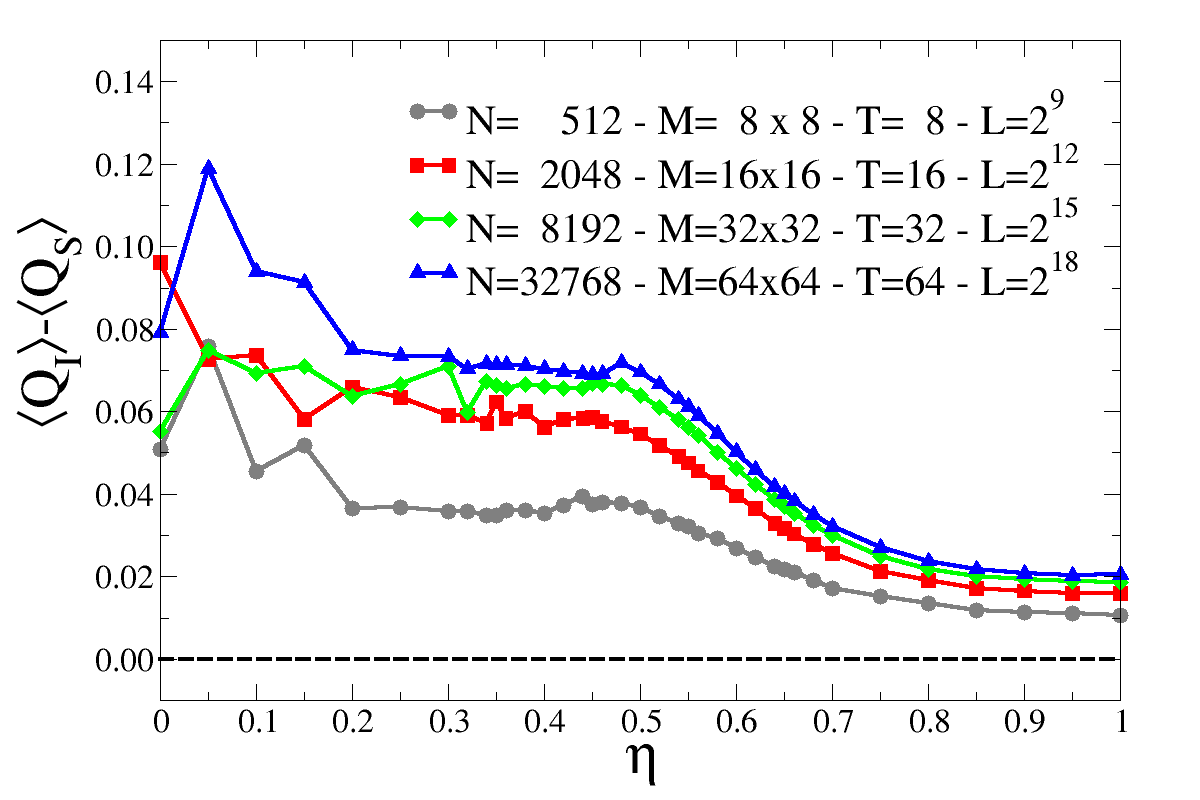} 
  \includegraphics[width=0.99\columnwidth]{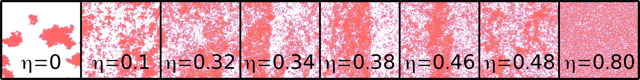} 
  \includegraphics[width=0.49\columnwidth]{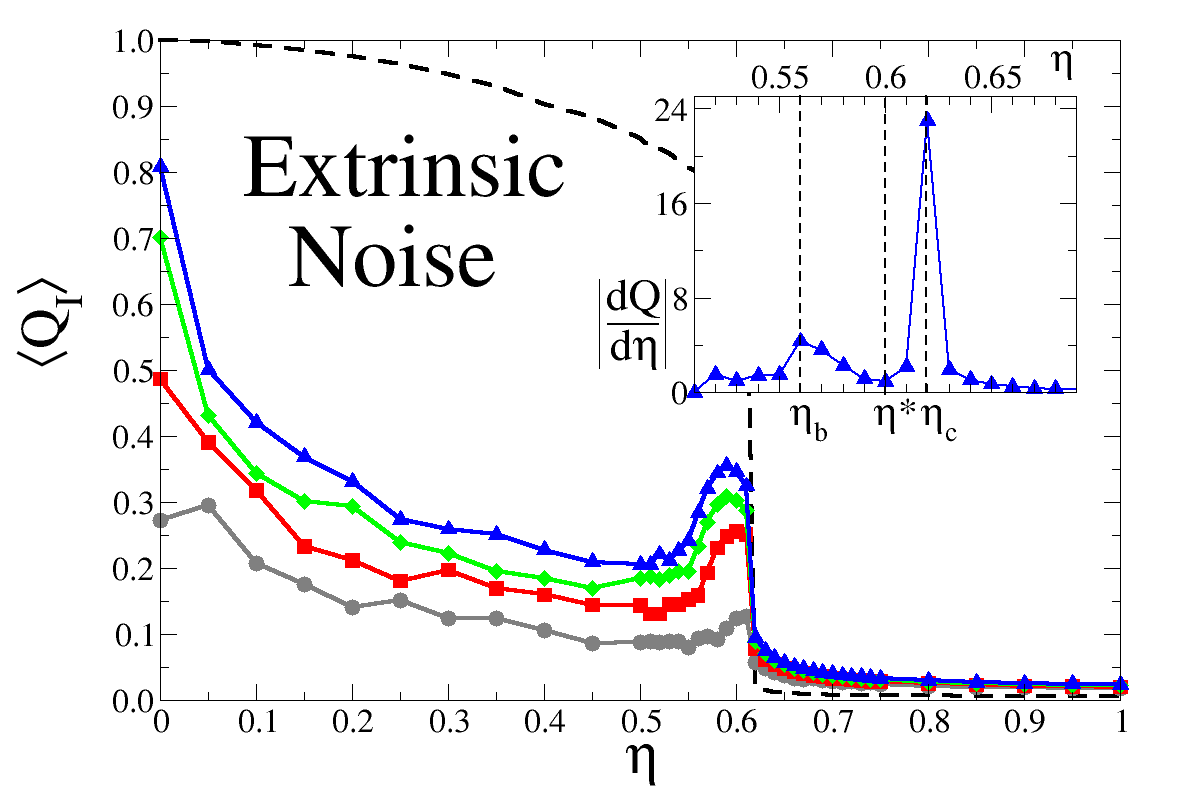} 
  \includegraphics[width=0.49\columnwidth]{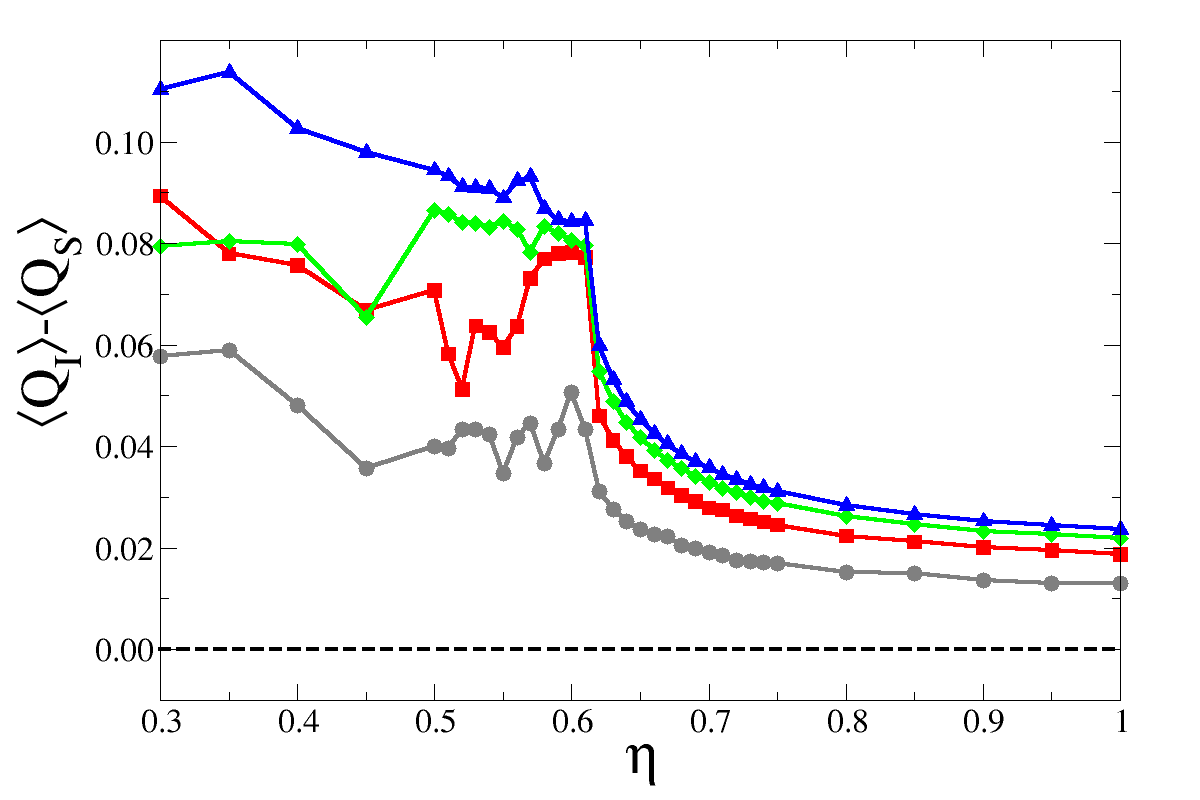} 
  \includegraphics[width=0.99\columnwidth]{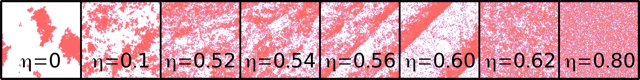}
  \caption{ Results obtained with VM simulations, both with intrinsic
    ({\bf Top}) and extrinsic ({\bf Bottom}) noise.\\ {\bf Left:}
    $\langle Q \rangle$ Vs $\eta$ computed with ITC scheme at
    increasing system size $N$ and time length $T$: dashed black lines
    refer to polarization $\Phi$ computed for $N=32768$. The inset
    shows the two peaks of $|dQ/d\eta|$ for $N=32768$ located at
    $\eta_b$ and $\eta_c$ (and the local maximum at $\eta^*$) which
    delimit the phase of traveling bands. The density structures are
    shown in the strips below.  We estimate $\eta_c\simeq 0.48$ and
    $\eta_b\simeq 0.34$ for intrinsic noise and $\eta_c\simeq0.62$ and
    $\eta_b\simeq 0.56$ for extrinsic one. \\ {\bf Right:} Differences
    with the STC scheme by varying the noise strength. }
  \label{fig:QVsETA}
\end{figure}

Fig.~\ref{fig:QVsETA} provides a detailed account of how $Q$ correlates
with the rich VM phase diagram.  Since we wish to
analyze the effect of system size $N$, and since the ITC
analysis requires a cubic matrix in space-time, we increase $T$ coherently with
$N$. 

We consider both variants of the VM, with intrinsic and extrinsic
noise, in order to investigate the sensitivity of  $Q$  to the known
differences between these two kinds of noise.  Intrinsic noise is
known to bear the signature of a smooth flocking transition
\cite{nagyvicsek}, while extrinsic noise exhibits a sharper
transition at a higher value of $\eta$ \cite{gregoire+al_04}. Both
behaviors are well reproduced by the $Q$ \textit{vs.} $\eta$ curve: in the
intrinsic noise case $Q$ has a smooth variation close to the known
value of $\eta_c$ (see also the dashed line reproducing the
polarization order parameter), while in the extrinsic noise case $Q$ has a
 rapid variation near $\eta_c$ (which is larger). 
 
 The variation of $Q$ in the vicinity of the transitions is made more clear by looking at the derivative
$|dQ/d\eta|$, shown in the insets on the left panels of Fig.~\ref{fig:QVsETA}. 
Remarkably, $Q$ does not signal only the polarization
transition, but also the other known crossover present in the VM
phenomenology. In particular, $Q$ reaches a local maximum at the point marked
by $\eta^*$.
The maximum decrease is marked by a second peak in
$|dQ/d\eta|$, at a noise value $\eta_b$.  This loss of order is well
explained by the behavior of the density field, represented in typical
configurations for selected values of $\eta$, in
Fig.~\ref{fig:QVsETA}. The polarization transition in the VM is
accompanied by traveling band structures. Such structures lose
coherence at smaller values of $\eta$, leading the density field to
become homogeneous. For even smaller values of $\eta<\eta_b$ we
observe a final increase of $Q$ with decreasing $\eta$. This is due to
the both the further increase of polarization and the appearance of
giant fluctuations, a well-studied phenomenon in the VM at low values of
noise \cite{ginelli}. Such strong inhomogeneities of the density field appear as
large areas with correlated values of the occupation field,
contributing to an increase of $Q$.

It is interesting to take another look at the difference between ITC
and STC.  In all cases, the the relative sensitivity of ITC with respect to STC
increases for larger $N$ and smaller $\eta$, with a few exceptions
(we note that at low values of noise the large correlations in the
system imply  large fluctuations).  In particular the difference
between STC and ITC increases close to the polarization
transition $\eta_c$ \cite{bagliettoalbano,gulish}.

\onecolumngrid

\twocolumngrid
\clearpage
\newpage

\subsection{Coping With Corrupted Data}
The VM is a numerical model that produces trajectories for which
identity, position, and velocity of each particle are exactly known at
each time step.  In real experiments on collective biological systems,
reconstructed trajectories (both in 2- and 3-dimensions) may become corrupted
in several possible ways, especially when analyzing large groups. 
For example, in some cases the trajectories of certain individuals are 
temporarily lost, so that their positions are
missing for several time steps.  Thus, some trajectories are
interrupted, meaning that at each time step we lack information about a
certain fraction of objects.  This uncertainty fraction typically grows with the system
density \cite{attanasi2015greta,sparta}.  

Here we examine whether our analysis is sensitive enough to show data on
ordering in the presence of data corruption, by
simulating the interruption of the trajectories by a simple two-state
Markov process. Such a process has two parameters that modulate the degree of data
corruption: the fraction of missing individuals $\mu$ and the
average length of a trajectory $\lambda$ (see Appendix for further details).
\begin{figure}[h!] 
  \includegraphics[width=\columnwidth]{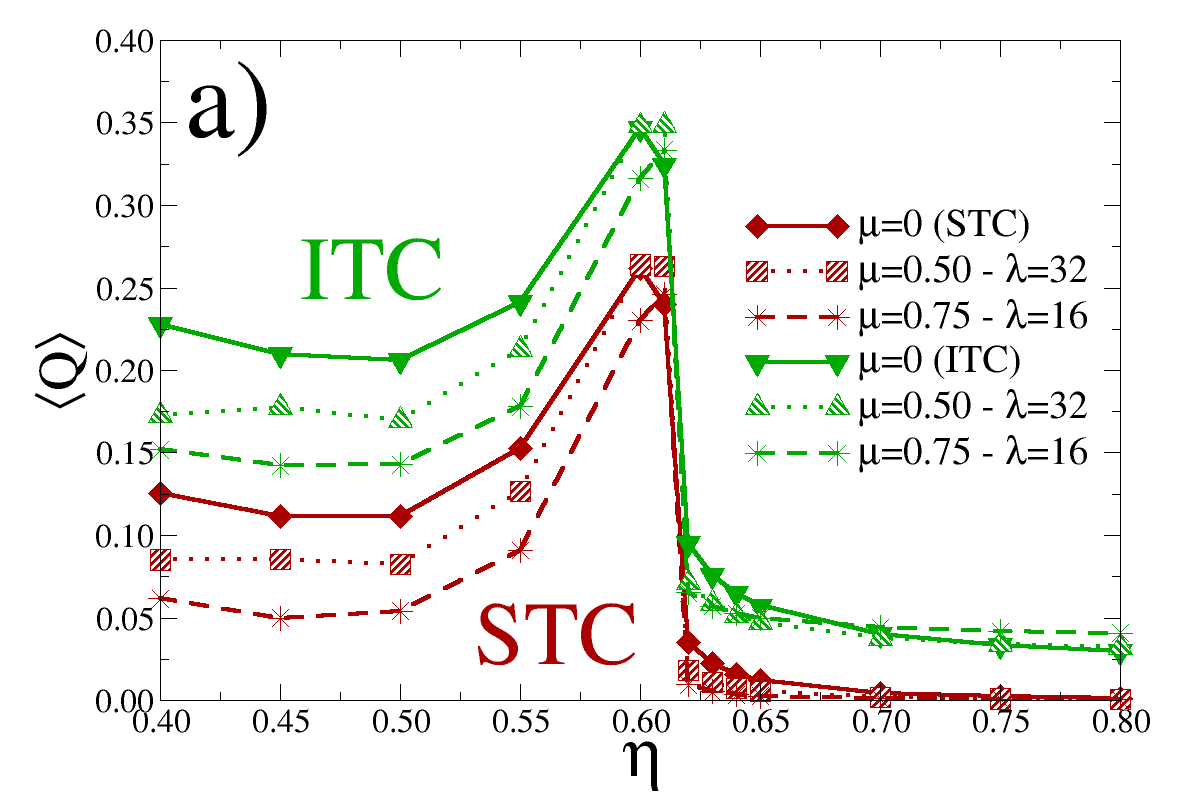} 
  \includegraphics[width=\columnwidth]{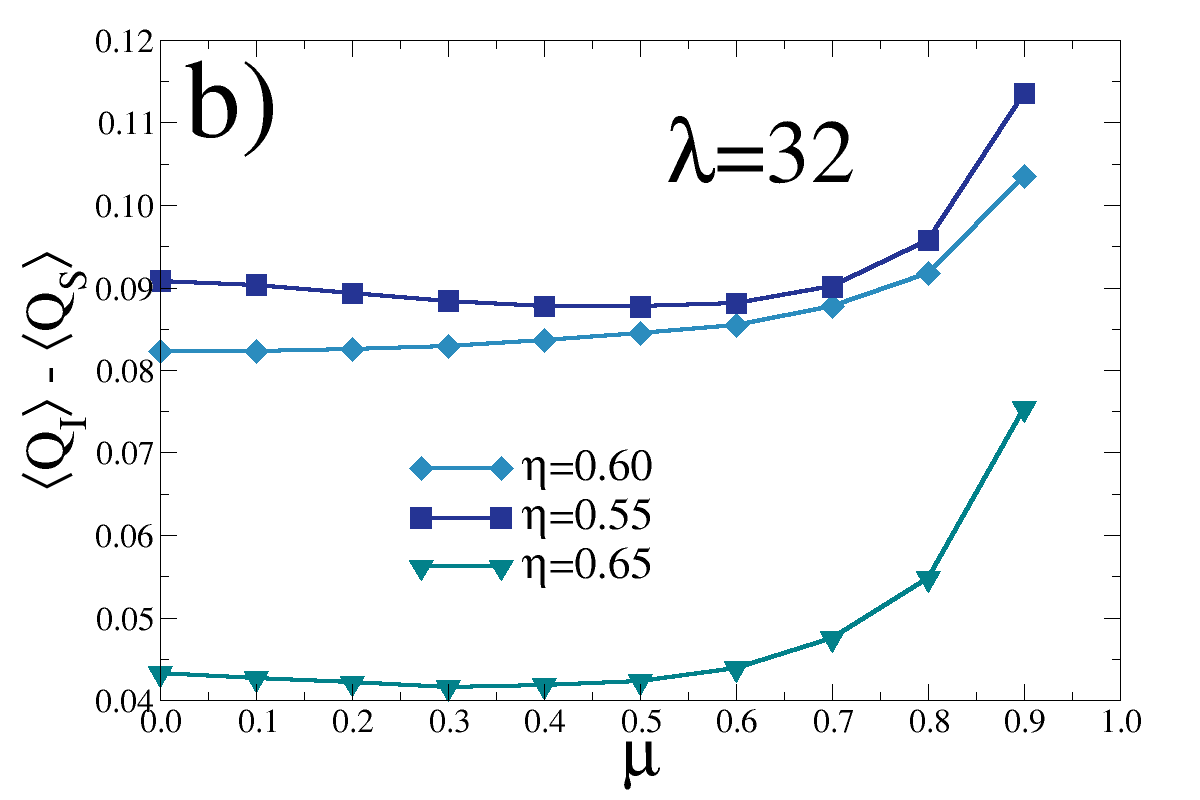} 
  \caption{{\bf Top} The effect of data corruption on $\langle Q \rangle$ {\it vs} $\eta$.  
  {\bf Bottom} Dependence on $\mu$ of the difference
    between $\langle Q_I \rangle$ and $\langle Q_S \rangle$ for some
    value of noise strength $\eta$. (VM simulations with extrinsic
    noise, parameters are $N=32768$, $L=64$ and $T=64$).}
 \label{fig:Degradation}
\end{figure}
 Fig. \ref{fig:Degradation} shows that both STC and ITC coding are
 able to detect the transition even in the presence of strong data
 corruption.  Since the corruption spoils correlations, the robustness
 of $Q$ is non-trivial. An important point is that, with increasing corruption of
 data, ITC copes with corruption significantly better than STC.  Not
 only $Q$ is always larger in the ITC case than in the STC one, but,
 more importantly, their difference {\it increases} with increasing
 data corruption (Fig. \ref{fig:Degradation}b), apart from few cases
 at small values of $\mu$.  This fact indicates that in real
 data-sets, ITC encoding will be more robust than STC. The reason
 seems to be that ITC exploits time correlations to recover the
 information which is lost because of disappeared particles.
\section{Conclusions}
We have studied a new definition of compression-based entropy and applied
it to the Vicsek model, an off-equilibrium active system which
describes collective behavior in biological systems.
We have adopted a crude encoding method, based upon a binary
coarse-graining of the positional information. We do not directly feed
the velocity information to the encoding, despite the crucial role of
velocity alignment in the VM.  Numerical results show that the
CID is able to capture the order-disorder
transition normally described by the velocity order parameter,
demonstrating that it can exploit the density-velocity coupling
present in the VM and in many non-equilibrium systems. This result is
promising for future analysis of real data, as the positional
information is easily obtained in experiments.  Of course, one could
in principle explore other, more refined, ways of encoding the
physical information, by changing, and expanding, the alphabet of the
compressed string according to other local properties of the system
(including velocities).  However, we have seen that a larger alphabet
makes the space of possible strings larger: therefore - at fixed
string length $L$ - it reduces the ability of the compressing scheme to
exploit correlations.  Given the results obtained with our simple
encoding method, we deem that more complex encodings are not
necessary in general.
Extending our approach to systems in three (or more) dimensions is
straightforward, since the Z-ordering curve has an obvious generalization in
any dimension.  The method is not constrained to regular cubic
lattices with $T=m=2^n$ (we have chosen to do so just to simplify
computational work).  In fact, regardless of the size of the
observation window and of grid features, once the cells have been
defined, it is sufficient to sort them according to their
Z-value. Effects of the aspect ratio of the simulation box is taken
into account by averaging on the permutations of the coordinates, as
illustrated in this paper.

Our results show that preserving locality in both space and time in the encoding (ITC),
rather then in space only (STC), is important. The ITC encoding always extracts more information than the
STC encoding and, most importantly, the difference in the performance
is robust (or even increases) in the presence of corruption of the data sets. This
result is particularly important if one wants to apply the method to
actual biological data.

We conclude by describing some future directions of our work.
The use of compression-based tools is particularly promising for
the study of the response to perturbations in collective biological
systems.  The fluctuation-dissipation theorem (FDT), connecting
the unperturbed correlations of a given observable to the linear
response of the system to a given small external perturbation, is
particularly simple at equilibrium, where the observables involved are
dictated by the Hamiltonian. In the case of flocks, swarms and other
biological systems one has to exploit one of the many recipes for
non-equilibrium generalization of FDT~\cite{BPRV08}. In all such recipes one needs
the knowledge of what are the relevant variables conjugated to the
perturbation; however, in the case of biological experiments, it is not
at clear what variables of the systems are perturbed in the presence of a
given external stimulus. Recent advances in non-equilibrium
statistical physics provide a possible way out: for non-equilibrium
steady states the response to perturbations can always be expressed in
terms of correlations involving observables conjugate with respect to
a specific observable, the `stochastic entropy' \cite{ss06}. We
conjecture that the observable $Q$ investigated here is closely related to
it. Our analysis, in particular, shows that $Q$ is coupled to many
relevant degrees of freedom in the system and therefore is a promising
candidate for a general approach to response in biological
systems. Future studies are needed to test this scenario.
\begin{acknowledgments}
We thank Irene Giardina and Stefania Melillo for the fruitful discussions.
This publication has been produced as a part of Italian-Israeli scientific cooperation project "IDA-CReBS":
A.C., A.P. and M.V. were supported by Italian Ministry of Foreign Affairs and International Cooperation and D.L. was supported by the Israel Ministry of Science and Technology (Grant no. 3-15150).
A.C. and M.V. were supported by European Research Council  (Advanced Grant  "RG.BIO" no. 785932).
A.P. acknowledges the financial support of Regione Lazio through the "Progetti Gruppi di Ricerca" (Grant  no. 85-2017-15257) and from the MIUR PRIN 2017 (Project no. 201798CZLJ).
D.L was supported by the Israel Science Foundation (Grant no. 1866/16) and the Israel-US Binational Science Foundation (Grant no. 2014713).
P.M.C, D.L., and S.M. were supported by the National Science Foundation Physics of Living Systems (Grant no. 1504867).
P.M.C. was partially supported by the Materials Research Science and Engineering Center Program of the National Science Foundation (Award no. DMR-1420073).
\end{acknowledgments} 
\section{Appendix}
\subsection{LZ77 Algorithm}
Let us illustrate how the LZ77 algorithm works, and derive the corresponding definition for the CID, with an example: we try to encode a sequence of characters in a list of ``longest previous factors'' (LPF). We represent a LPF by pair of integer $(p,l)$ which are the ``instructions'' to retrieve the original sequence: {\it``print $l$ symbols starting from $p$-th character already written, if $l=0$ just print $p$ directly''}.
To decode we must go through the list of LPFs in order, and for each of them follow the instructions.
Let's consider the following binary sequence of length $L=16$
$$
0101100111001110
$$
and read it from left to right.
At the beginning no characters have been printed yet so, surely, the first two LPFs will be $(\text{`}0\text{'},0)$ and $(\text{`}1\text{'},0)$ 
(both have $l=0$, in this case $p$ is the character to be printed, we emphasize it using quotation marks).
Next we have a substring `01' that we have already met: the LPF is then $(1,2)$ (copy $2$ characters starting from position $1$).
The next substring already met is `10' and it is encoded by LPF $(2,2)$, followed by $(3,3)$ to encode `011'.
Finally, we see that the remaining substring `1001110' is obtained by copying $7$ characters starting from position $5$, so the last LSF is $(5,7)$.
It does not matter if the sequence currently available is shorter than $7$ characters and incomplete, as the full subsequence will become available during printing. The number of bits $\mathcal{L}$ needed to encode this list of $C=6$ LPFs can be estimated by the following argument.
Let $(p_i,l_i)$ be the $i$-th LPF and $b_i \simeq \log{p_i}+ \log{l_i}$ the number of bits needed to encode it (all $\log$s are base $2$):
in this way the length of the original sequence $L$ and the compressed binary length $\mathcal{L}$ are given by:
$$
L \simeq \sum_{i=1}^C l_i \qquad \mathcal{L}= \sum_{i=1}^{C} b_i
$$
Since $p_i<L$ we have $b_i\leq(\log{L}+\log{l_i})$, $\mathcal{L}$ must be bounded by
$$
\mathcal{L}  \leq C\log{L} + \sum_{i=1}^C\log{l_i}
$$
Now, we use Jensen's inequality for concave functions
$$
\sum_{i=1}^C \log{l_i} \leq C \log{\left(\frac{1}{C}\sum_{i=1}^{C} l_i\right)}=C\log\frac{L}{C}
$$
and, after simple algebraic manipulation, we obtain our definition of CID as in \cite{martiniani2019quantifying}
$$
\frac{\mathcal{L}}{L} \leq \frac{C\log{C} + 2C\log{(L/C)}}{L} \equiv \CID
$$
The CID of our example ($C=6$, $L=16$) is $2.03$. The sequence is too short and we doubled the length of the original sequence.
For effective compression we must consider longer sequences, for example consider an $L=128$ string obtained by replicating the sequence considered above ($L=16$) 8 times. In this case we must add one more LPF, $(1,112)$, so, since $C=7$ and $L=128$, we find $\CID\simeq 0.61$ and we have almost halved the number of bits needed to represent the sequence.
\subsection{Data Corruption}
In order to mimic real data corruption, or degradation, we proceed as follows.
We  associate to each particle $i$  a boolean random variable $b_i\in\{0,1\}$ which evolves under the action of a 2-states Markov Chain with transition probability $P(b \to b')=P_{bb'}$ :
$$
p = P_{10} = 1-P_{11} \qquad
q = P_{01} = 1-P_{00}
$$
At each simulation step we apply the rules of this Markov Chain to evolve stochastically the $b_i$.
As a result, we are able to modulate the degree of data corruption by tuning two parameters: {\it i)} the fraction of missing individuals $\mu$ and, {\it ii)} the average length of a trajectory, $\lambda$.
In detail, we consider or ignore particle $i$ of the data-set according to the value of $b_i$: when $b_i=0$ the particle $i$ is removed from the data-set until $b_i$ returns to $1$. It easy to prove that the invariant measure $\rho_b$ ($\rho_0=\mu$) and the typical length (the average life-span) $\lambda$ of a trajectory depend on $p$ and $q$ in the following way:
\begin{eqnarray*}
\mu &=& \rho_0 = 1 - \rho_1 = \frac{p}{p+q}\\
\lambda &=& \frac{\sum_{l=1}^{\infty} l P_{11}^l}{\sum_{l=0}^{\infty} P_{11}^l}=\frac{1-p}{p}
\end{eqnarray*}
Then, by setting $p$ and $q$ to appropriate values 
\begin{equation*}
p = \frac{1}{ 1 + \lambda} \qquad 
q = \frac{1-\mu}{\mu} p
\end{equation*}
and by starting with a configuration of $\{b_i\}_{i=1,N}$ already in equilibrium according to the invariant measure $\sum_i b_i / N \simeq \rho_1$, we can simulate data corruption by varying $\mu$ and $\lambda$.
%


\begin{thebibliography}{36}%
\makeatletter
\providecommand \@ifxundefined [1]{%
 \@ifx{#1\undefined}
}%
\providecommand \@ifnum [1]{%
 \ifnum #1\expandafter \@firstoftwo
 \else \expandafter \@secondoftwo
 \fi
}%
\providecommand \@ifx [1]{%
 \ifx #1\expandafter \@firstoftwo
 \else \expandafter \@secondoftwo
 \fi
}%
\providecommand \natexlab [1]{#1}%
\providecommand \enquote  [1]{``#1''}%
\providecommand \bibnamefont  [1]{#1}%
\providecommand \bibfnamefont [1]{#1}%
\providecommand \citenamefont [1]{#1}%
\providecommand \href@noop [0]{\@secondoftwo}%
\providecommand \href [0]{\begingroup \@sanitize@url \@href}%
\providecommand \@href[1]{\@@startlink{#1}\@@href}%
\providecommand \@@href[1]{\endgroup#1\@@endlink}%
\providecommand \@sanitize@url [0]{\catcode `\\12\catcode `\$12\catcode
  `\&12\catcode `\#12\catcode `\^12\catcode `\_12\catcode `\%12\relax}%
\providecommand \@@startlink[1]{}%
\providecommand \@@endlink[0]{}%
\providecommand \url  [0]{\begingroup\@sanitize@url \@url }%
\providecommand \@url [1]{\endgroup\@href {#1}{\urlprefix }}%
\providecommand \urlprefix  [0]{URL }%
\providecommand \Eprint [0]{\href }%
\providecommand \doibase [0]{http://dx.doi.org/}%
\providecommand \selectlanguage [0]{\@gobble}%
\providecommand \bibinfo  [0]{\@secondoftwo}%
\providecommand \bibfield  [0]{\@secondoftwo}%
\providecommand \translation [1]{[#1]}%
\providecommand \BibitemOpen [0]{}%
\providecommand \bibitemStop [0]{}%
\providecommand \bibitemNoStop [0]{.\EOS\space}%
\providecommand \EOS [0]{\spacefactor3000\relax}%
\providecommand \BibitemShut  [1]{\csname bibitem#1\endcsname}%
\let\auto@bib@innerbib\@empty
\bibitem [{\citenamefont {Zurek}(2018)}]{zurek2018complexity}%
  \BibitemOpen
  \bibfield  {author} {\bibinfo {author} {\bibfnamefont {W.~H.}\ \bibnamefont
  {Zurek}},\ }\href@noop {} {\emph {\bibinfo {title} {Complexity, entropy and
  the physics of information}}}\ (\bibinfo  {publisher} {CRC Press},\ \bibinfo
  {year} {2018})\BibitemShut {NoStop}%
\bibitem [{\citenamefont {Chaitin}(1990)}]{chaitin1990information}%
  \BibitemOpen
  \bibfield  {author} {\bibinfo {author} {\bibfnamefont {G.~J.}\ \bibnamefont
  {Chaitin}},\ }\href@noop {} {\emph {\bibinfo {title} {Information, randomness
  \& incompleteness: papers on algorithmic information theory}}},\
  Vol.~\bibinfo {volume} {8}\ (\bibinfo  {publisher} {World Scientific},\
  \bibinfo {year} {1990})\BibitemShut {NoStop}%
\bibitem [{\citenamefont {Mezard}\ and\ \citenamefont
  {Montanari}(2009)}]{mezard2009information}%
  \BibitemOpen
  \bibfield  {author} {\bibinfo {author} {\bibfnamefont {M.}~\bibnamefont
  {Mezard}}\ and\ \bibinfo {author} {\bibfnamefont {A.}~\bibnamefont
  {Montanari}},\ }\href@noop {} {\emph {\bibinfo {title} {Information, physics,
  and computation}}}\ (\bibinfo  {publisher} {Oxford University Press},\
  \bibinfo {year} {2009})\BibitemShut {NoStop}%
\bibitem [{\citenamefont {Marconi}\ \emph {et~al.}(2008)\citenamefont
  {Marconi}, \citenamefont {Puglisi}, \citenamefont {Rondoni},\ and\
  \citenamefont {Vulpiani}}]{BPRV08}%
  \BibitemOpen
  \bibfield  {author} {\bibinfo {author} {\bibfnamefont {U.~M.~B.}\
  \bibnamefont {Marconi}}, \bibinfo {author} {\bibfnamefont {A.}~\bibnamefont
  {Puglisi}}, \bibinfo {author} {\bibfnamefont {L.}~\bibnamefont {Rondoni}}, \
  and\ \bibinfo {author} {\bibfnamefont {A.}~\bibnamefont {Vulpiani}},\
  }\href@noop {} {\bibfield  {journal} {\bibinfo  {journal} {Phys. Rep.}\
  }\textbf {\bibinfo {volume} {461}},\ \bibinfo {pages} {111} (\bibinfo {year}
  {2008})}\BibitemShut {NoStop}%
\bibitem [{\citenamefont {{Shannon}}(1948)}]{shannon}%
  \BibitemOpen
  \bibfield  {author} {\bibinfo {author} {\bibfnamefont {C.~E.}\ \bibnamefont
  {{Shannon}}},\ }\href {\doibase 10.1002/j.1538-7305.1948.tb00917.x}
  {\bibfield  {journal} {\bibinfo  {journal} {The Bell System Technical
  Journal}\ }\textbf {\bibinfo {volume} {27}},\ \bibinfo {pages} {623}
  (\bibinfo {year} {1948})}\BibitemShut {NoStop}%
\bibitem [{\citenamefont {Vogel}\ \emph {et~al.}(2009)\citenamefont {Vogel},
  \citenamefont {Saravia}, \citenamefont {Bachmann}, \citenamefont {Fierro},\
  and\ \citenamefont {Fischer}}]{vogel2009phase}%
  \BibitemOpen
  \bibfield  {author} {\bibinfo {author} {\bibfnamefont {E.}~\bibnamefont
  {Vogel}}, \bibinfo {author} {\bibfnamefont {G.}~\bibnamefont {Saravia}},
  \bibinfo {author} {\bibfnamefont {F.}~\bibnamefont {Bachmann}}, \bibinfo
  {author} {\bibfnamefont {B.}~\bibnamefont {Fierro}}, \ and\ \bibinfo {author}
  {\bibfnamefont {J.}~\bibnamefont {Fischer}},\ }\href@noop {} {\bibfield
  {journal} {\bibinfo  {journal} {Physica A: Statistical Mechanics and its
  Applications}\ }\textbf {\bibinfo {volume} {388}},\ \bibinfo {pages} {4075}
  (\bibinfo {year} {2009})}\BibitemShut {NoStop}%
\bibitem [{\citenamefont {Melchert}\ and\ \citenamefont
  {Hartmann}(2015)}]{melchert}%
  \BibitemOpen
  \bibfield  {author} {\bibinfo {author} {\bibfnamefont {O.}~\bibnamefont
  {Melchert}}\ and\ \bibinfo {author} {\bibfnamefont {A.}~\bibnamefont
  {Hartmann}},\ }\href@noop {} {\bibfield  {journal} {\bibinfo  {journal}
  {Phys. Rev. E}\ }\textbf {\bibinfo {volume} {91}},\ \bibinfo {pages} {023306}
  (\bibinfo {year} {2015})}\BibitemShut {NoStop}%
\bibitem [{\citenamefont {Lebowitz}\ and\ \citenamefont {Spohn}(1999)}]{LS99}%
  \BibitemOpen
  \bibfield  {author} {\bibinfo {author} {\bibfnamefont {J.~L.}\ \bibnamefont
  {Lebowitz}}\ and\ \bibinfo {author} {\bibfnamefont {H.}~\bibnamefont
  {Spohn}},\ }\href@noop {} {\bibfield  {journal} {\bibinfo  {journal} {J.
  Stat. Phys.}\ }\textbf {\bibinfo {volume} {95}},\ \bibinfo {pages} {333}
  (\bibinfo {year} {1999})}\BibitemShut {NoStop}%
\bibitem [{\citenamefont {Parrondo}\ \emph {et~al.}(2009)\citenamefont
  {Parrondo}, \citenamefont {Van~den Broeck},\ and\ \citenamefont
  {Kawai}}]{parrondo2009}%
  \BibitemOpen
  \bibfield  {author} {\bibinfo {author} {\bibfnamefont {J.~M.}\ \bibnamefont
  {Parrondo}}, \bibinfo {author} {\bibfnamefont {C.}~\bibnamefont {Van~den
  Broeck}}, \ and\ \bibinfo {author} {\bibfnamefont {R.}~\bibnamefont
  {Kawai}},\ }\href@noop {} {\bibfield  {journal} {\bibinfo  {journal} {New J.
  Phys.}\ }\textbf {\bibinfo {volume} {11}},\ \bibinfo {pages} {073008}
  (\bibinfo {year} {2009})}\BibitemShut {NoStop}%
\bibitem [{\citenamefont {Martiniani}\ \emph {et~al.}(2019)\citenamefont
  {Martiniani}, \citenamefont {Chaikin},\ and\ \citenamefont
  {Levine}}]{martiniani2019quantifying}%
  \BibitemOpen
  \bibfield  {author} {\bibinfo {author} {\bibfnamefont {S.}~\bibnamefont
  {Martiniani}}, \bibinfo {author} {\bibfnamefont {P.~M.}\ \bibnamefont
  {Chaikin}}, \ and\ \bibinfo {author} {\bibfnamefont {D.}~\bibnamefont
  {Levine}},\ }\href@noop {} {\bibfield  {journal} {\bibinfo  {journal} {Phys.
  Rev. X}\ }\textbf {\bibinfo {volume} {9}},\ \bibinfo {pages} {011031}
  (\bibinfo {year} {2019})}\BibitemShut {NoStop}%
\bibitem [{\citenamefont {Martiniani}\ \emph {et~al.}(2020)\citenamefont
  {Martiniani}, \citenamefont {Lemberg}, \citenamefont {Chaikin},\ and\
  \citenamefont {Levine}}]{martiniani2020correlation}%
  \BibitemOpen
  \bibfield  {author} {\bibinfo {author} {\bibfnamefont {S.}~\bibnamefont
  {Martiniani}}, \bibinfo {author} {\bibfnamefont {Y.}~\bibnamefont {Lemberg}},
  \bibinfo {author} {\bibfnamefont {P.~M.}\ \bibnamefont {Chaikin}}, \ and\
  \bibinfo {author} {\bibfnamefont {D.}~\bibnamefont {Levine}},\ }\href@noop {}
  {\bibfield  {journal} {\bibinfo  {journal} {arXiv preprint arXiv:2004.03502}\
  } (\bibinfo {year} {2020})}\BibitemShut {NoStop}%
\bibitem [{\citenamefont {Benedetto}\ \emph {et~al.}(2002)\citenamefont
  {Benedetto}, \citenamefont {Caglioti},\ and\ \citenamefont
  {Loreto}}]{loreto}%
  \BibitemOpen
  \bibfield  {author} {\bibinfo {author} {\bibfnamefont {D.}~\bibnamefont
  {Benedetto}}, \bibinfo {author} {\bibfnamefont {E.}~\bibnamefont {Caglioti}},
  \ and\ \bibinfo {author} {\bibfnamefont {V.}~\bibnamefont {Loreto}},\
  }\href@noop {} {\bibfield  {journal} {\bibinfo  {journal} {Phys. Rev. Lett.}\
  }\textbf {\bibinfo {volume} {88}},\ \bibinfo {pages} {048702} (\bibinfo
  {year} {2002})}\BibitemShut {NoStop}%
\bibitem [{\citenamefont {Puglisi}\ \emph {et~al.}(2003)\citenamefont
  {Puglisi}, \citenamefont {Benedetto}, \citenamefont {Caglioti}, \citenamefont
  {Loreto},\ and\ \citenamefont {Vulpiani}}]{puglisi2003data}%
  \BibitemOpen
  \bibfield  {author} {\bibinfo {author} {\bibfnamefont {A.}~\bibnamefont
  {Puglisi}}, \bibinfo {author} {\bibfnamefont {D.}~\bibnamefont {Benedetto}},
  \bibinfo {author} {\bibfnamefont {E.}~\bibnamefont {Caglioti}}, \bibinfo
  {author} {\bibfnamefont {V.}~\bibnamefont {Loreto}}, \ and\ \bibinfo {author}
  {\bibfnamefont {A.}~\bibnamefont {Vulpiani}},\ }\href@noop {} {\bibfield
  {journal} {\bibinfo  {journal} {Physica D}\ }\textbf {\bibinfo {volume}
  {180}},\ \bibinfo {pages} {92} (\bibinfo {year} {2003})}\BibitemShut
  {NoStop}%
\bibitem [{\citenamefont {Henkel}\ \emph {et~al.}(2008)\citenamefont {Henkel},
  \citenamefont {Hinrichsen},\ and\ \citenamefont
  {L{\"u}beck}}]{Non-Equilibrium_Book}%
  \BibitemOpen
  \bibfield  {author} {\bibinfo {author} {\bibfnamefont {M.}~\bibnamefont
  {Henkel}}, \bibinfo {author} {\bibfnamefont {H.}~\bibnamefont {Hinrichsen}},
  \ and\ \bibinfo {author} {\bibfnamefont {S.}~\bibnamefont {L{\"u}beck}},\
  }\href@noop {} {\emph {\bibinfo {title} {Non-Equilibrium Phase Transitions -
  Volume 1: Absorbing Phase Transitions}}}\ (\bibinfo  {publisher} {Springer},\
  \bibinfo {year} {2008})\BibitemShut {NoStop}%
\bibitem [{\citenamefont {Cates}\ and\ \citenamefont
  {Tailleur}(2015)}]{cates15}%
  \BibitemOpen
  \bibfield  {author} {\bibinfo {author} {\bibfnamefont {M.~E.}\ \bibnamefont
  {Cates}}\ and\ \bibinfo {author} {\bibfnamefont {J.}~\bibnamefont
  {Tailleur}},\ }\href@noop {} {\bibfield  {journal} {\bibinfo  {journal}
  {Annu. Rev. Condens. Matter Phys.}\ }\textbf {\bibinfo {volume} {6}},\
  \bibinfo {pages} {219} (\bibinfo {year} {2015})}\BibitemShut {NoStop}%
\bibitem [{\citenamefont {Cover}\ and\ \citenamefont
  {Thomas}(2012)}]{cover2012elements}%
  \BibitemOpen
  \bibfield  {author} {\bibinfo {author} {\bibfnamefont {T.~M.}\ \bibnamefont
  {Cover}}\ and\ \bibinfo {author} {\bibfnamefont {J.~A.}\ \bibnamefont
  {Thomas}},\ }\href@noop {} {\emph {\bibinfo {title} {Elements of information
  theory}}}\ (\bibinfo  {publisher} {John Wiley \& Sons},\ \bibinfo {year}
  {2012})\BibitemShut {NoStop}%
\bibitem [{\citenamefont {Sheinwald}\ \emph {et~al.}(1990)\citenamefont
  {Sheinwald}, \citenamefont {Lempel},\ and\ \citenamefont
  {Ziv}}]{sheinwald1990two}%
  \BibitemOpen
  \bibfield  {author} {\bibinfo {author} {\bibfnamefont {D.}~\bibnamefont
  {Sheinwald}}, \bibinfo {author} {\bibfnamefont {A.}~\bibnamefont {Lempel}}, \
  and\ \bibinfo {author} {\bibfnamefont {J.}~\bibnamefont {Ziv}},\ }\href@noop
  {} {\bibfield  {journal} {\bibinfo  {journal} {IEEE Trans. Commun.}\ }\textbf
  {\bibinfo {volume} {38}},\ \bibinfo {pages} {341} (\bibinfo {year}
  {1990})}\BibitemShut {NoStop}%
\bibitem [{\citenamefont {Vicsek}\ \emph {et~al.}(1995)\citenamefont {Vicsek},
  \citenamefont {Czir{\'o}k}, \citenamefont {Ben-Jacob}, \citenamefont
  {Cohen},\ and\ \citenamefont {Shochet}}]{vicsek+al_95}%
  \BibitemOpen
  \bibfield  {author} {\bibinfo {author} {\bibfnamefont {T.}~\bibnamefont
  {Vicsek}}, \bibinfo {author} {\bibfnamefont {A.}~\bibnamefont {Czir{\'o}k}},
  \bibinfo {author} {\bibfnamefont {E.}~\bibnamefont {Ben-Jacob}}, \bibinfo
  {author} {\bibfnamefont {I.}~\bibnamefont {Cohen}}, \ and\ \bibinfo {author}
  {\bibfnamefont {O.}~\bibnamefont {Shochet}},\ }\href@noop {} {\bibfield
  {journal} {\bibinfo  {journal} {Phys Rev Lett}\ }\textbf {\bibinfo {volume}
  {75}},\ \bibinfo {pages} {1226} (\bibinfo {year} {1995})}\BibitemShut
  {NoStop}%
\bibitem [{\citenamefont {Parrish}\ and\ \citenamefont
  {Hamner}(1997)}]{parrish_review}%
  \BibitemOpen
  \bibfield  {author} {\bibinfo {author} {\bibfnamefont {J.~K.}\ \bibnamefont
  {Parrish}}\ and\ \bibinfo {author} {\bibfnamefont {W.~M.}\ \bibnamefont
  {Hamner}},\ }\href@noop {} {\emph {\bibinfo {title} {Animal Groups in Three
  Dimensions}}}\ (\bibinfo  {publisher} {Cambridge University Press},\ \bibinfo
  {address} {Cambridge},\ \bibinfo {year} {1997})\BibitemShut {NoStop}%
\bibitem [{\citenamefont {Gr{\'e}goire}\ \emph {et~al.}(2003)\citenamefont
  {Gr{\'e}goire}, \citenamefont {Chat{\'e}},\ and\ \citenamefont
  {Tu}}]{gregoire2003moving}%
  \BibitemOpen
  \bibfield  {author} {\bibinfo {author} {\bibfnamefont {G.}~\bibnamefont
  {Gr{\'e}goire}}, \bibinfo {author} {\bibfnamefont {H.}~\bibnamefont
  {Chat{\'e}}}, \ and\ \bibinfo {author} {\bibfnamefont {Y.}~\bibnamefont
  {Tu}},\ }\href@noop {} {\bibfield  {journal} {\bibinfo  {journal} {Physica D:
  Nonlinear Phenomena}\ }\textbf {\bibinfo {volume} {181}},\ \bibinfo {pages}
  {157} (\bibinfo {year} {2003})}\BibitemShut {NoStop}%
\bibitem [{\citenamefont {Aldana}\ \emph {et~al.}(2007)\citenamefont {Aldana},
  \citenamefont {Dossetti}, \citenamefont {Huepe}, \citenamefont {Kenkre},\
  and\ \citenamefont {Larralde}}]{aldana2007phase}%
  \BibitemOpen
  \bibfield  {author} {\bibinfo {author} {\bibfnamefont {M.}~\bibnamefont
  {Aldana}}, \bibinfo {author} {\bibfnamefont {V.}~\bibnamefont {Dossetti}},
  \bibinfo {author} {\bibfnamefont {C.}~\bibnamefont {Huepe}}, \bibinfo
  {author} {\bibfnamefont {V.}~\bibnamefont {Kenkre}}, \ and\ \bibinfo {author}
  {\bibfnamefont {H.}~\bibnamefont {Larralde}},\ }\href@noop {} {\bibfield
  {journal} {\bibinfo  {journal} {Physical review letters}\ }\textbf {\bibinfo
  {volume} {98}},\ \bibinfo {pages} {095702} (\bibinfo {year}
  {2007})}\BibitemShut {NoStop}%
\bibitem [{\citenamefont {Baglietto}\ and\ \citenamefont
  {Albano}(2009)}]{baglietto2009nature}%
  \BibitemOpen
  \bibfield  {author} {\bibinfo {author} {\bibfnamefont {G.}~\bibnamefont
  {Baglietto}}\ and\ \bibinfo {author} {\bibfnamefont {E.~V.}\ \bibnamefont
  {Albano}},\ }\href@noop {} {\bibfield  {journal} {\bibinfo  {journal}
  {Physical Review E}\ }\textbf {\bibinfo {volume} {80}},\ \bibinfo {pages}
  {050103} (\bibinfo {year} {2009})}\BibitemShut {NoStop}%
\bibitem [{Note1()}]{Note1}%
  \BibitemOpen
  \bibinfo {note} {Hence this is a {\protect \it metric} implementation of
  VM}\BibitemShut {NoStop}%
\bibitem [{\citenamefont {Solon}\ \emph {et~al.}(2015)\citenamefont {Solon},
  \citenamefont {Chat\'e},\ and\ \citenamefont {Tailleur}}]{solon}%
  \BibitemOpen
  \bibfield  {author} {\bibinfo {author} {\bibfnamefont {A.~P.}\ \bibnamefont
  {Solon}}, \bibinfo {author} {\bibfnamefont {H.}~\bibnamefont {Chat\'e}}, \
  and\ \bibinfo {author} {\bibfnamefont {J.}~\bibnamefont {Tailleur}},\ }\href
  {\doibase 10.1103/PhysRevLett.114.068101} {\bibfield  {journal} {\bibinfo
  {journal} {Phys. Rev. Lett.}\ }\textbf {\bibinfo {volume} {114}},\ \bibinfo
  {pages} {068101} (\bibinfo {year} {2015})}\BibitemShut {NoStop}%
\bibitem [{\citenamefont {Ginelli}(2016)}]{ginelli}%
  \BibitemOpen
  \bibfield  {author} {\bibinfo {author} {\bibfnamefont {F.}~\bibnamefont
  {Ginelli}},\ }\href@noop {} {\bibfield  {journal} {\bibinfo  {journal} {The
  European Physical Journal Special Topics}\ }\textbf {\bibinfo {volume}
  {225}},\ \bibinfo {pages} {2099} (\bibinfo {year} {2016})}\BibitemShut
  {NoStop}%
\bibitem [{\citenamefont {Gr{\'e}goire}\ and\ \citenamefont
  {Chat{\'e}}(2004)}]{gregoire+al_04}%
  \BibitemOpen
  \bibfield  {author} {\bibinfo {author} {\bibfnamefont {G.}~\bibnamefont
  {Gr{\'e}goire}}\ and\ \bibinfo {author} {\bibfnamefont {H.}~\bibnamefont
  {Chat{\'e}}},\ }\href@noop {} {\bibfield  {journal} {\bibinfo  {journal}
  {Phys Rev Lett}\ }\textbf {\bibinfo {volume} {92}},\ \bibinfo {pages}
  {025702} (\bibinfo {year} {2004})}\BibitemShut {NoStop}%
\bibitem [{\citenamefont {Chat{\'e}}\ \emph {et~al.}(2008)\citenamefont
  {Chat{\'e}}, \citenamefont {Ginelli}, \citenamefont {Gr{\'e}goire},\ and\
  \citenamefont {Raynaud}}]{chate+al_08b}%
  \BibitemOpen
  \bibfield  {author} {\bibinfo {author} {\bibfnamefont {H.}~\bibnamefont
  {Chat{\'e}}}, \bibinfo {author} {\bibfnamefont {F.}~\bibnamefont {Ginelli}},
  \bibinfo {author} {\bibfnamefont {G.}~\bibnamefont {Gr{\'e}goire}}, \ and\
  \bibinfo {author} {\bibfnamefont {F.}~\bibnamefont {Raynaud}},\ }\href@noop
  {} {\bibfield  {journal} {\bibinfo  {journal} {Phys Rev E Stat Nonlin Soft
  Matter Phys}\ }\textbf {\bibinfo {volume} {77}},\ \bibinfo {pages} {046113}
  (\bibinfo {year} {2008})}\BibitemShut {NoStop}%
\bibitem [{\citenamefont {Máté~Nagy}(2007)}]{nagyvicsek}%
  \BibitemOpen
  \bibfield  {author} {\bibinfo {author} {\bibfnamefont {T.~V.}\ \bibnamefont
  {Máté~Nagy}, \bibfnamefont {István~Daruka}},\ }\href@noop {} {\bibfield
  {journal} {\bibinfo  {journal} {Physica A: Statistical Mechanics and its
  Applications}\ }\textbf {\bibinfo {volume} {373}},\ \bibinfo {pages} {445}
  (\bibinfo {year} {2007})}\BibitemShut {NoStop}%
\bibitem [{\citenamefont {Aldana}\ \emph {et~al.}(2009)\citenamefont {Aldana},
  \citenamefont {Larralde},\ and\ \citenamefont {V\'{a}zquez}}]{aldanadebate}%
  \BibitemOpen
  \bibfield  {author} {\bibinfo {author} {\bibfnamefont {M.}~\bibnamefont
  {Aldana}}, \bibinfo {author} {\bibfnamefont {H.}~\bibnamefont {Larralde}}, \
  and\ \bibinfo {author} {\bibfnamefont {B.}~\bibnamefont {V\'{a}zquez}},\
  }\href@noop {} {\bibfield  {journal} {\bibinfo  {journal} {International
  Journal of Modern Physics B}\ }\textbf {\bibinfo {volume} {23}},\ \bibinfo
  {pages} {3661} (\bibinfo {year} {2009})}\BibitemShut {NoStop}%
\bibitem [{\citenamefont {Morton}(1966)}]{morton1966computer}%
  \BibitemOpen
  \bibfield  {author} {\bibinfo {author} {\bibfnamefont {G.~M.}\ \bibnamefont
  {Morton}},\ }\href@noop {} {\  (\bibinfo {year} {1966})}\BibitemShut
  {NoStop}%
\bibitem [{Note2()}]{Note2}%
  \BibitemOpen
  \bibinfo {note} {In our case, the entries take on the values $0$ or $1$, but
  this is not important for the scan.}\BibitemShut {Stop}%
\bibitem [{\citenamefont {Baglietto}\ and\ \citenamefont
  {Albano}(2006)}]{bagliettoalbano}%
  \BibitemOpen
  \bibfield  {author} {\bibinfo {author} {\bibfnamefont {G.}~\bibnamefont
  {Baglietto}}\ and\ \bibinfo {author} {\bibfnamefont {E.~V.}\ \bibnamefont
  {Albano}},\ }\href {\doibase 10.1142/S0129183106008492} {\bibfield  {journal}
  {\bibinfo  {journal} {International Journal of Modern Physics C}\ }\textbf
  {\bibinfo {volume} {17}},\ \bibinfo {pages} {395} (\bibinfo {year}
  {2006})}\BibitemShut {NoStop}%
\bibitem [{\citenamefont {Gulich}\ \emph {et~al.}(2018)\citenamefont {Gulich},
  \citenamefont {Baglietto},\ and\ \citenamefont {Rozenfeld}}]{gulish}%
  \BibitemOpen
  \bibfield  {author} {\bibinfo {author} {\bibfnamefont {D.}~\bibnamefont
  {Gulich}}, \bibinfo {author} {\bibfnamefont {G.}~\bibnamefont {Baglietto}}, \
  and\ \bibinfo {author} {\bibfnamefont {A.~F.}\ \bibnamefont {Rozenfeld}},\
  }\href {\doibase https://doi.org/10.1016/j.physa.2018.02.094} {\bibfield
  {journal} {\bibinfo  {journal} {Physica A: Statistical Mechanics and its
  Applications}\ }\textbf {\bibinfo {volume} {502}},\ \bibinfo {pages} {590 }
  (\bibinfo {year} {2018})}\BibitemShut {NoStop}%
\bibitem [{\citenamefont {Attanasi}\ \emph {et~al.}(2015)\citenamefont
  {Attanasi}, \citenamefont {Cavagna}, \citenamefont {Del~Castello},
  \citenamefont {Giardina}, \citenamefont {Jeli{\'c}}, \citenamefont {Melillo},
  \citenamefont {Parisi}, \citenamefont {Pellacini}, \citenamefont {Shen},
  \citenamefont {Silvestri} \emph {et~al.}}]{attanasi2015greta}%
  \BibitemOpen
  \bibfield  {author} {\bibinfo {author} {\bibfnamefont {A.}~\bibnamefont
  {Attanasi}}, \bibinfo {author} {\bibfnamefont {A.}~\bibnamefont {Cavagna}},
  \bibinfo {author} {\bibfnamefont {L.}~\bibnamefont {Del~Castello}}, \bibinfo
  {author} {\bibfnamefont {I.}~\bibnamefont {Giardina}}, \bibinfo {author}
  {\bibfnamefont {A.}~\bibnamefont {Jeli{\'c}}}, \bibinfo {author}
  {\bibfnamefont {S.}~\bibnamefont {Melillo}}, \bibinfo {author} {\bibfnamefont
  {L.}~\bibnamefont {Parisi}}, \bibinfo {author} {\bibfnamefont
  {F.}~\bibnamefont {Pellacini}}, \bibinfo {author} {\bibfnamefont
  {E.}~\bibnamefont {Shen}}, \bibinfo {author} {\bibfnamefont {E.}~\bibnamefont
  {Silvestri}},  \emph {et~al.},\ }\href@noop {} {\bibfield  {journal}
  {\bibinfo  {journal} {IEEE transactions on pattern analysis and machine
  intelligence}\ }\textbf {\bibinfo {volume} {37}},\ \bibinfo {pages} {2451}
  (\bibinfo {year} {2015})}\BibitemShut {NoStop}%
\bibitem [{\citenamefont {{Cavagna}}\ \emph {et~al.}(2019)\citenamefont
  {{Cavagna}}, \citenamefont {{Melillo}}, \citenamefont {{Parisi}},\ and\
  \citenamefont {{Ricci-Tersenghi}}}]{sparta}%
  \BibitemOpen
  \bibfield  {author} {\bibinfo {author} {\bibfnamefont {A.}~\bibnamefont
  {{Cavagna}}}, \bibinfo {author} {\bibfnamefont {S.}~\bibnamefont
  {{Melillo}}}, \bibinfo {author} {\bibfnamefont {L.}~\bibnamefont {{Parisi}}},
  \ and\ \bibinfo {author} {\bibfnamefont {F.}~\bibnamefont
  {{Ricci-Tersenghi}}},\ }\href {\doibase 10.1109/TPAMI.2019.2946796}
  {\bibfield  {journal} {\bibinfo  {journal} {IEEE Transactions on Pattern
  Analysis and Machine Intelligence}\ ,\ \bibinfo {pages}
  {doi:10.1109/TPAMI.2019.2946796}} (\bibinfo {year} {2019})}\BibitemShut
  {NoStop}%
\bibitem [{\citenamefont {Speck}\ and\ \citenamefont {Seifert}(2006)}]{ss06}%
  \BibitemOpen
  \bibfield  {author} {\bibinfo {author} {\bibfnamefont {T.}~\bibnamefont
  {Speck}}\ and\ \bibinfo {author} {\bibfnamefont {U.}~\bibnamefont
  {Seifert}},\ }\href@noop {} {\bibfield  {journal} {\bibinfo  {journal}
  {Europhys. Lett.}\ }\textbf {\bibinfo {volume} {74}},\ \bibinfo {pages} {391}
  (\bibinfo {year} {2006})}\BibitemShut {NoStop}%
\end{thebibliography}

\end{document}